\documentclass{elsart}

\usepackage{natbib}
\usepackage{subfigure}
\usepackage{rotate}

\usepackage{epsfig}

\usepackage{amssymb}

\journal{Journal of Theoretical Biology}
\begin{document}
\newfont{\tc}{tcrm1200}

\begin{frontmatter}


  \title{Active Brownian Particle and Random Walk Theories of the Motions of
    Zooplankton: Application to Experiments with Swarms of {\em Daphnia}
                                %\thanksref{dedic}
    }
  %\thanks[dedic]{This work is dedicated to \ldots}
\author[hub]{Udo Erdmann\corauthref{cor1}},
\ead{udo.erdmann@physik.hu-berlin.de}
\ead[url]{http://www.udoerdmann.de/}
\corauth[cor1]{Corresponding author.}
  \author[hub]{Werner Ebeling},
  \author[hub]{Lutz Schimansky-Geier},
  \author[umsl]{Anke Ordemann},
  \author[umsl]{Frank Moss}
  \address[hub]{Institute of Physics, Humboldt-University Berlin, 12489
    Berlin, Newtonstr. 15, Germany}
  \address[umsl]{Center for Neurodynamics, University of Missouri at St.
    Louis, St. Louis, MO 63121, USA}

\begin{abstract}
% Text of abstract
  Active Brownian Particles are self-propelled particles that move in a
  dissipative medium subject to random forces, or ``noise''. Additionally,
  they can be confined by an external field and/or they can interact with one
  another. The external field may actually be an attractive marker, for
  example a light field (as in the experiment) or an energy potential or a
  chemical gradient (as in the theory). The potential energy can also be the
  result of interparticle attractive and/or repulsive forces summed over all
  particles (a mean field potential). Four, qualitatively different motions of
  the particles are possible: at small particle density their motions are
  approximately independent of one another subject only to the external field
  and the noise, which results in moving randomly through or performing
  rotational motions about a central point in space. At increasing densities
  interactions play an important role and individuals form a swarm performing
  several types of self-organized collective motion. We apply this model for
  the description of zooplankton {\em Daphnia} swarms. In the case of the
  zooplankton {\em Daphnia} (and probably many other aquatic animals that form
  similar motions as well) this vortex is hydrodynamical but motivated by the
  self-propelled motion of the individuals. Similar vortex-type motions have
  been observed for other creatures ranging in size from bacteria to flocks of
  birds and schools of fish. However, our experiment with {\em Daphnia} is
  unique in that all four motions can be observed in controlled laboratory
  conditions with the same animal. Moreover, the theory, presented in both
  continuous differential equation and random walk forms, offers a
  quantitative, physically based explanation of the four motions.
\end{abstract}


\end{frontmatter}

% main text
\section{Introduction}
The theory of Active Brownian Particles (ABP) has been developed in some
detail over the past several years and published largely in physics literature
\citep{SchiMiRoMa95,Er99,SchwEbTi98,EbSchwTi99,ErEbSchiSchw99,Schw03}. The ABP
theory is based on continuous motions in time and is described by a set of
stochastic differential equations. A second approach has been developed, which
is based on the random walk theory (called RWT from now on). It has been
applied to particles that move in discrete ``hops'' with a pause at the end of
each hop \citep{OrBaCaMo03,OrBaMo02a,OrBaMo02b}. Though originally conceived
without a self-propelling velocity \citep{SchiMiRoMa95}, the present ABP
theory deals with particles that are self-propelled and move through a
dissipative medium, such as water. The particles expend energy in order to
move and to supply their own metabolic requirements, while foraging for food
represents an energy uptake.

These notions are conveniently expressed in terms of an effective, velocity
dependent dissipation, $\gamma(v)$, which can assume positive and for
small velocities possibly negative values. For $\gamma(v) < 0$, energy
is pumped into the particle and may be stored in a depot \citep{EbSchwTi99}
(there is more than sufficient food to supply the energies of motion and
metabolism), while for $\gamma(v) > 0$ the rate of energy uptake and
conversion is insufficient to maintain the supply in the depot causing the
particle to slow down. The particle then finds a velocity for which, on
average, the effective dissipation is zero, and this value is the
self-propelling velocity. See Fig.~\ref{fig:1} below. The right/left symmetry
of these motions may be broken by parallelizing interactions, what could be
any kind of a velocity aligning force. Single or groups of ABPs can move in a
central potential which insures that they are confined to some region in
space. Here we confine our model to a quadratic potential representing a
linear restoring force. In addition, a random force, or noise, is applied in
order to represent the variability inherent in a real animal's motions.  In
contrast to the ABP theory, the RWT begins with the assumption of an average,
constant self-propelling velocity, as does another approach
\citep{ViCzBeCoSh95}. Each particle moves by hopping and turning. At the end
of each hop a change in direction is taken from an experimentally measured
turning angle distribution and a ``kick'' toward a center of motion is
applied.  Random selections from the turning angle distribution represent the
noise and the kick insures that the particles are confined. Here we take the
kick to be a force linearly proportional to the distance between the particle
and the center of the field.  Both theories describe systems that operate far
from thermodynamic equilibrium, since motion is maintained in the dissipative
medium by a continual uptake of energy.

In the ABP theory, coherent motions at low density are modeled by interactions
with the central field plus noise only. Two attractors motions are separated
by a bifurcation controlled by the magnitude of the self-propelling velocity
\citep{SchwEbTi98}: one a noisy fixed point and the other a pair of symmetric,
limit cycles (recurrent closed paths) with equal probabilities to show a
clockwise (CW) or counterclockwise (CCW) rotational sense in a plane
\citep{ErEbSchiSchw99}. See Fig. 2 below. Other systems, for example molecular
motors \citep{JuAiPr97}, which are also far from equilibrium and behave
according to a similar velocity dependent dissipation, also show bimodal
velocity distributions, that is, motion in two opposing directions that is
equally probable \citep{BaJuPr02}. The theory presented here is
two-dimensional so that these motions are confined to the $x$-$y$ plane. A
particle-particle attractive interaction is then built in
\citep{EbSchw01a,ErEbAn00}. The interaction is global in the sense that every
particle feels the attraction of every other one but the strength decreases
with pair separation distance. Such interactions result in a central mean
field and predict swarming behavior, the gradual increase of particle density
around the symmetry axis of the central field. Finally, a particle-particle
hydrodynamic interaction is included that models the tendency for nearby
particles (feeling one another's flow field) to align the directions of their
velocities \citep{ErEb02}. This interaction breaks the symmetry of the two
limit cycles leading to a collective vortex-type motion in one or the other
rotational direction. A short range repulsive potential that results in
avoidance when two particles come closer than some minimum distance can also
be used to break the symmetry and lead to vortex formation. The sense of
rotation of the vortex depends on an initial condition that is random, so that
the final motion can be either CCW or CW.

Thus the ABP theory predicts four basic types of motion if the particles are
confined: 1) noisy fixed point whereby a particle moves randomly about and
through a single point in space; 2) two limit cycles whereby the particle
moves on closed paths with CCW or CW directions equally probable. The second
type can be separated again in two different types of motion if we investigate
larger populations of interacting individuals: 3.1) swarming whereby initially
randomly and widely dispersed particles first move toward a spatial center and
then start rotating around this center (rotational motions in both directions
are equally probable, and thus the net rotational motion is on average zero);
and 3.2) vortex whereby all particles spontaneously align their velocities and
commence rotation in the same direction, because of a higher density of the
individuals. As we show below, all four of these motions can be observed in
laboratory experiments within a single species: the light sensitive
zooplankton {\em Daphnia}.

Such motions, in addition to other collective behaviors, are commonly observed
\citep{PaEd99} in schools of fish \citep{We73a,We73b,HaWaMa86,PaViGr02},
bacterial colonies \citep{BeShoTeCoCziVi94}, slime molds
\citep{LeRe91,Bo98,DaKeEn00} and flocks of birds \citep{LaFr88}. Vortex
motions were specifically observed long ago in colonies of ants by Schneirla
\citep{ArToRoLe72} and more recently in bacterial colonies \citep{CziBeCoVi96}
and slime molds \citep{RaNiSaLe99}.  Swarming is of course quite well known.
See for example \citep{OkLe02,PaViGr02}. The idea of a self-propelling
velocity arose early \citep{We73b}, and was expressed by Sakai whose model
predicted the aforementioned four types of motion already in 1973.  See a
discussion on that early work in the useful book \citep{OkLe02}. General
reviews on the physics of individual particle motions and aggregation
(swarming) has been given by \citet{BeCoLe00} and \citet{FlGrLeOl99}. More
recent theories, based on the idea of a self-propelling velocity, incorporated
various couplings among the particles such as directional averaging of the
vector velocities \citep{BeShoTeCoCziVi94,ViCzBeCoSh95,LeRaCo00,CoKrJaRuFr02}.
A hydrodynamic generalization of these ``rule based'' theories was later
developed by \citet{ToTu95,ToTu98}. In contrast to these, the ABP theory,
based on the energy depot model \citep{EbSchwTi99}, is represented by
stochastic differential equations.

Our objective here is to summarize the ABP and RWT theories without burdening
the reader with excessive detail and to compare the predictions of those
theories at each step in their development with our experimental results on
the motions of {\em Daphnia}.

\section{Outline of the ABP Theory: single particle case}
\label{sec:outline-abp-theory}

In 1827 the British botanist, Robert Brown, observed irregular motions of
microscopic particles in water. He noted that their motions resembled that of
living creatures. Later all such entities whose motions are mediated by random
forces came to be called ``Brownian Particles''. In a two-dimensional spatial
plane, $x$-$y$, they are located by a vector $\vec{r}$, whose magnitude is
$r=\sqrt{x^2+y^2}$ and move with velocity, $\partial_t \vec{r}=\vec{v}$.
Physical models represented by stochastic differential equations, called
Langevin equations, describe the motion of these particles. They represent a
balance of forces on the particle: the random force, or noise $\sqrt{2D}
\vec{\xi}(t)$, where $D$ is the noise strength; the confining restoring force,
$\nabla U(r)$, where $U(r)$ is the external potential energy, the passive
friction $-\gamma_0 \vec{v}$ and, in our case, the self-propelling force $d\,
e(t) \vec{v}$ that results from conversion of energy out of an internal energy
depot $e(t)$ into motion;
\begin{equation}
  \label{eq:1}
  m \partial_t \vec{v}=-\gamma_0 \vec{v}+ d\, e(t) \vec{v}-\nabla U(r)
                       +\sqrt{2D}\vec{\xi}(t)\,.
\end{equation}
The noise, $\vec{\xi}(t)$, in this equation is external in the sense that it
is contributed by the environment in which the particle finds itself. In the
case of ABPs this noise could arise, for example, from thermal collisions,
hydrodynamic fluctuations or turbulence. This noise is assumed to be
``white'', that is its autocorrelation function is a delta function, $\langle
\xi_i(t) \xi_j(s)\rangle=2D\delta_{ij}\,\delta(t-s)$, Gaussian distributed and
with zero mean.  As an identifier, we will call this the ``external'' noise.
Further, from now on we will choose units in which $m \equiv 1$. The time
evolution of energy stored in the depot is given by $e(t)$, which is described
by a second equation,
\begin{equation}
  \label{eq:2}
  \partial_t e(t)=q_0-c e(t) -d\, v^2\, e(t)\,,
\end{equation}
that represents a balance among three processes: $q_0$ which is the energy
inflow available to the depot, here called the uptake, for example from
foraging for food; $-c e(t)$, which is a constant drain representing, for
example, a metabolic requirement; and $d\,v^2\e(t)$, which is the power
dissipated necessary to propel the particle through the medium. Here, $c$ and
$d$ are constants. We take the available energy inflow $q_0$, to be uniformly
distributed over the two-dimensional space, though in more general treatments
it can be irregularly distributed, $q(\vec{r})$ representing, for example,
food availability in patches \citep{SteEbCa94,SchwEbTi98}.

\begin{figure}[htbp]
  \begin{center}
    \epsfig{file=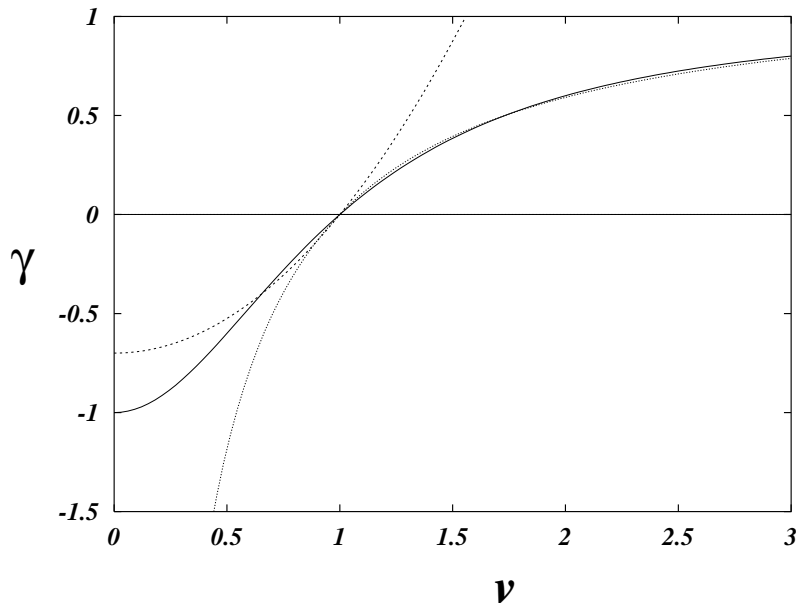,width=11cm}
    \caption{The velocity dependent friction function, Eq.~(\ref{eq:3})
      showing sub- and super-critical energy pumping regions separated by the
      self-propelling velocity for which the effective dissipation is zero.
      The Rayleigh and Gruler formulae are shown by the dotted and dashed
      lines, respectively. The values of the parameters are, $q_0 =10$, $c =
      1.0$, $\gamma_0 = 2.0$, and $d = 0.7$}
    \label{fig:1}
  \end{center}
\end{figure}

\begin{figure}[htbp]
  \begin{center}
    \subfigure[$\mu = 0.35$]{\label{fig:2a}
      \epsfig{file=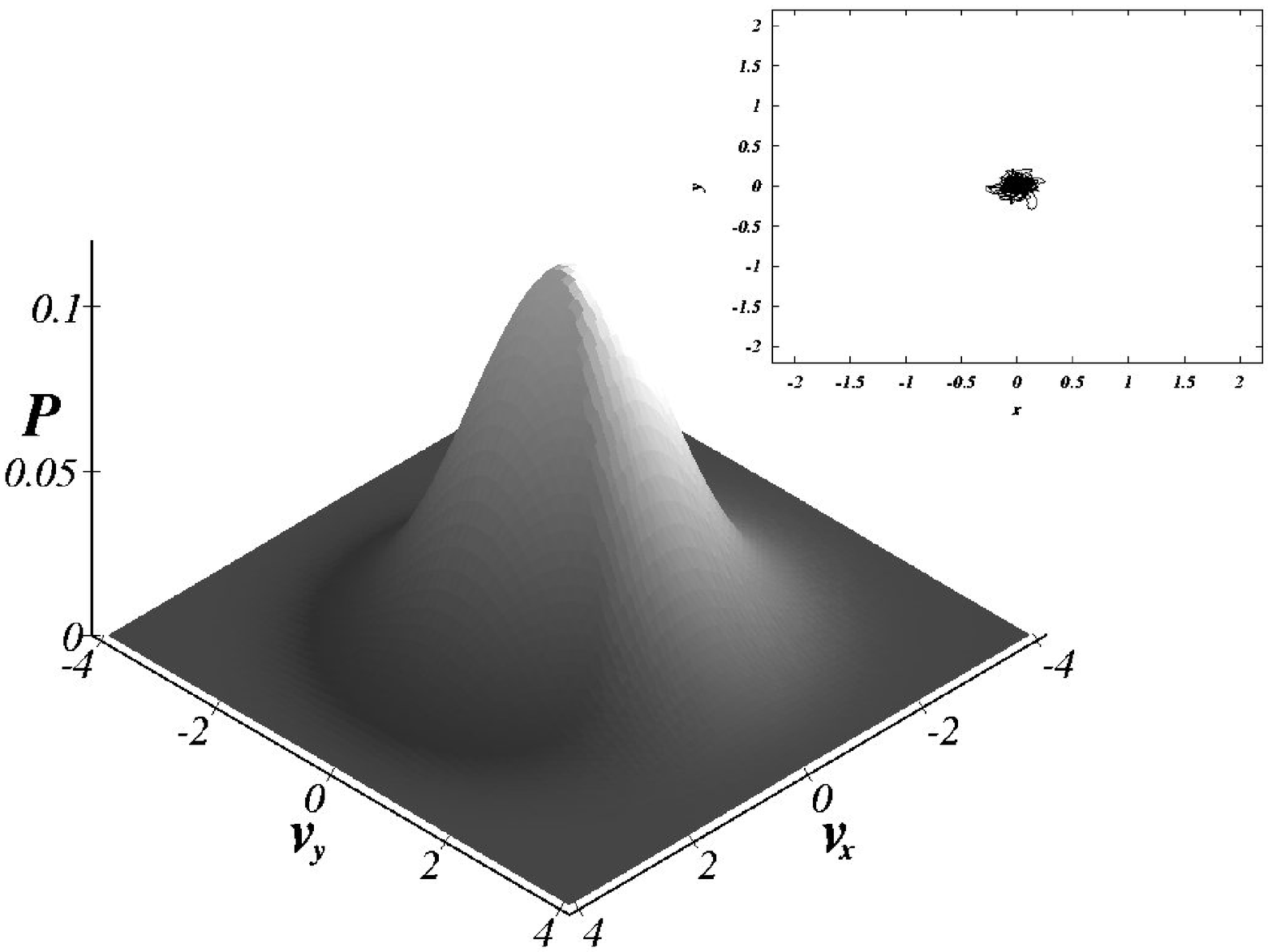,width=6.7cm}}
    \subfigure[$\mu = 3.5$]{\label{fig:2b}
      \epsfig{file=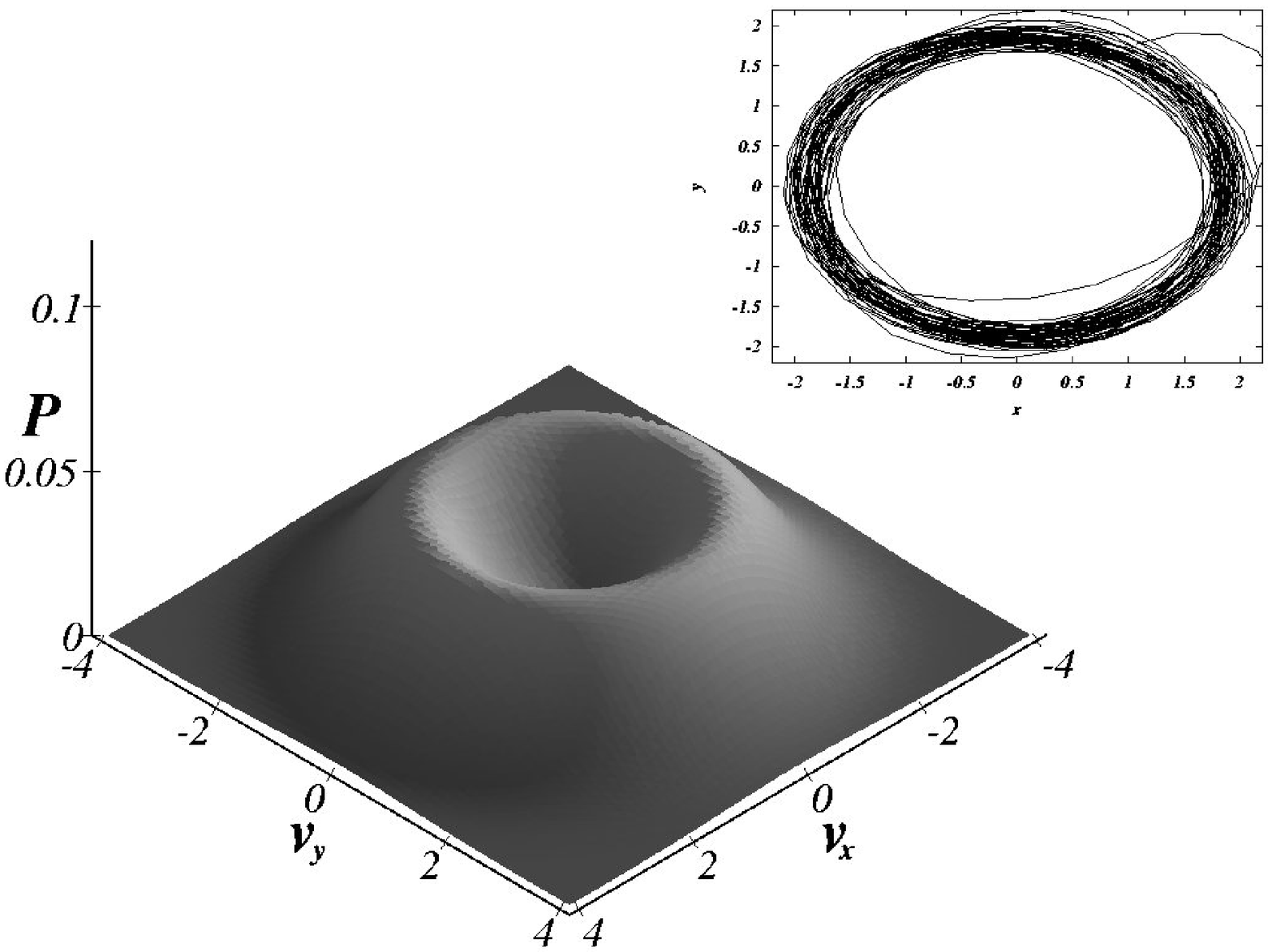,width=6.7cm}}
    \caption{The probability densities of velocity in the two-dimensional 
      plane showing the noisy fixed point for $\mu = 0.35$ in (a) and the
      limit cycles for $\mu = 3.5$ in (b).  The other parameters are $\gamma_0
      = 2.0$, $D = 2.0$, $c = 1.0$ and $q_0 = 10$.  The insets show typical
      trajectories in the plane}
    \label{fig:2}
  \end{center}
\end{figure}

A key issue is the velocity dependent dissipation, $\gamma(v)$. Here we
mention two different commonly used treatments.  The first is the Rayleigh
law, $\gamma(v)=-\gamma_1+\gamma_2 v^2$, where the two $\gamma$'s are
constants serving to delineate constant and velocity dependent contributions
to the dissipation. After initial transients die out, the particle assumes a
self-propelling velocity, $v_0^2=\gamma_1/\gamma_2$. This law was originally
developed to describe dissipative energy pumping in systems involving sound
\citep{Ra94engl} and has been used to describe self-propelling velocities
applied to fish schooling \citep{Ni94,Ni96a}. The second is a law developed by
\citet{SchiGr93}, $\gamma(v)=\gamma_0(1-v_0/v)$, developed to describe the
motions of various motile biological cells. The same friction law was used
independently by \citet{HeMo95} to describe the motion of pedestrians. Both of
these laws have singularities, the former in a range of $v \gg v_0$, and the
later for $v \rightarrow 0$. A key simplification can be obtained by assuming
that the energy uptake rate $e(t)$ is slow compared with the dissipative
rates.  Equation~(\ref{eq:2}) can then be adiabatically eliminated,
$\partial_t e(t)=0$, which, together with Eq.~(\ref{eq:1}), results in a third
dissipation law that does not suffer from singularities and thus is more
physically reasonable \citep{ErEbSchiSchw99},
\begin{equation}
  \label{eq:3}
  \gamma(v)=\gamma_0-\frac{q_0 d}{c+dv^2}\,,
\end{equation}
where $\gamma_0$ is the dissipation constant in the limit of high particle
velocity. This function is shown in Fig.~\ref{fig:1} where the velocity ranges
of energy pumping into the depot (super-critical pumping) and dissipation from
it (sub-critical pumping) are indicated separated by the self-propelling
velocity at zero effective dissipation,
\begin{equation}
  \label{eq:4}
  v_0^2=\frac{q_0}{\gamma_0}-\frac cd\,.
\end{equation}
The solutions of Eq.~(\ref{eq:1}) using the formula Eq.~(\ref{eq:3}) give the
first two types of motion separated by a bifurcation. The bifurcation
parameter can be defined as the ratio of constants,
\begin{equation}
  \label{eq:5}
  \mu=\frac{q_0 d}{c \gamma_0}\,.
\end{equation}
Introducing a force $\nabla U(\vec{r})=\omega_0^2\vec{r}$, attraction by a
light shaft, chemotaxis and similar causes can be modeled. For such a
quadratic attractive potential (the force is linear in $\vec{r}$)
two-dimensional motion with natural frequency $\omega_0$ on closed paths in
the $x$-$y$ plane is possible. For $\mu < 1$, there is sub-critical pumping of
energy and the motion is random around the symmetry point of the central
potential, called here the ``noisy fixed point''; and for $\mu > 1$ the
pumping is super-critical, and the resulting motion is a symmetric pair of
noisy limit cycles. Particles move on the limit cycles with the
self-propelling velocity $|\vec{v}|>0$. These limit cycles are closed curves
in the $x$-$y$ plane, and their symmetry represents the fact that the two
senses of rotation, CCW and CW, occur with equal probability. This bifurcation
is shown in Fig.~\ref{fig:2}, where the probability densities of the velocity
components $v_x$ and $v_y$ are plotted showing the monomodal peak (noisy fixed
point) in Fig.~\ref{fig:2a} for $\mu < 1$ and the bimodal density (limit
cycles) in Fig.~\ref{fig:2b} for $\mu > 1$. The insets show example
trajectories in the $x$-$y$ plane.

\begin{figure}[htbp]
  \begin{center}
    \epsfig{file=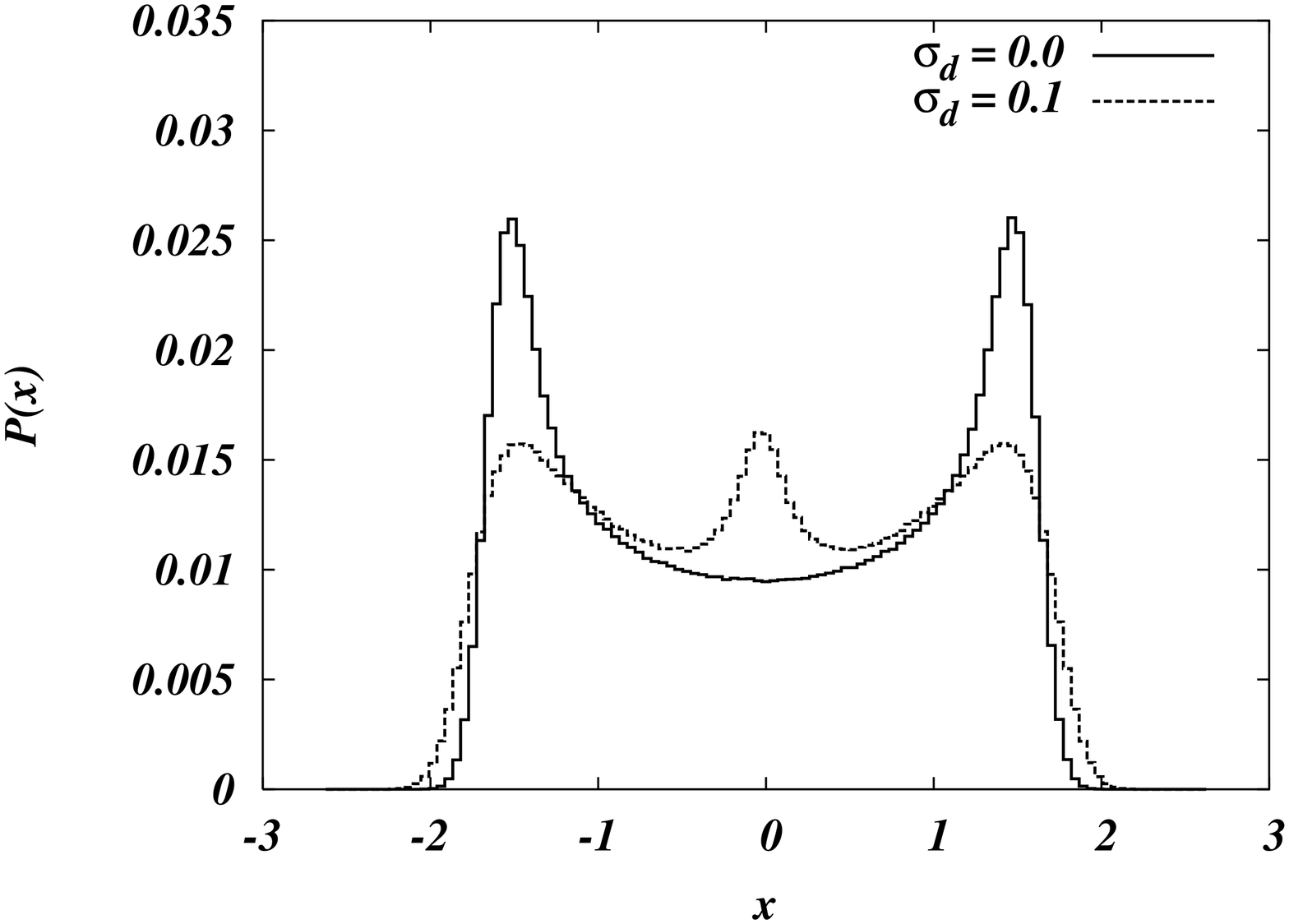,height=6.8cm,angle=-90}
    \epsfig{file=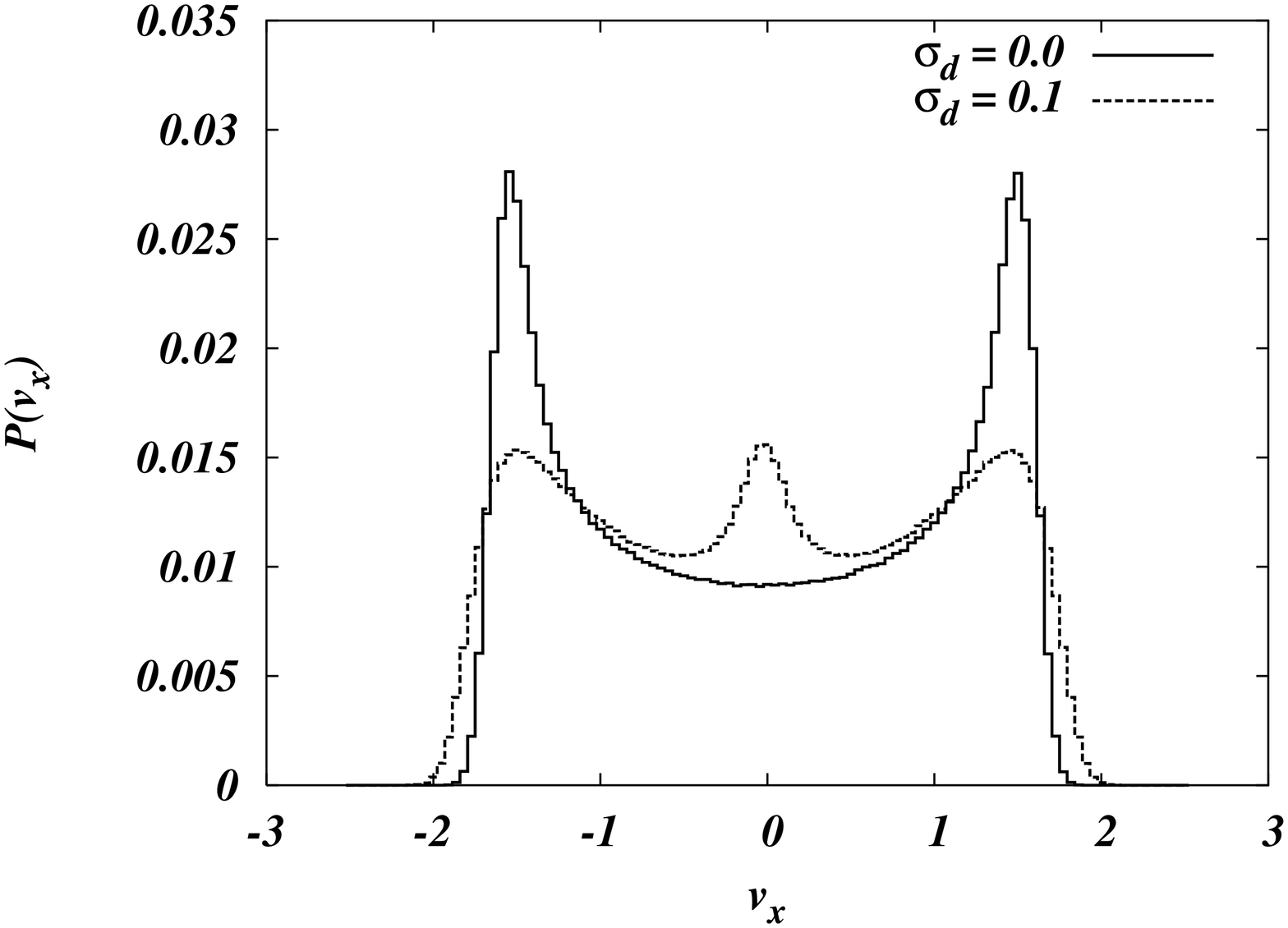,height=6.8cm,angle=-90}
    \caption{The probability density of radii and velocities 
      projected on the $x$-axis for the single noise $\vec{\xi}(t)$ in
      Eqs.~(\ref{eq:1}) and (\ref{eq:2}) shown by the solid line, and for
      both noises, $\vec{\xi}(t)$ and $\eta(t)$, in Eqs.~(\ref{eq:1}) and
      (\ref{eq:6}) shown by the dashed curve.}
    \label{fig:3}
  \end{center}
\end{figure}

The foregoing theory applies to a single particle or to a collection of
identical non-interacting particles moving at a unique self-propelling
velocity in a attractive central potential. Below, however, we compare the
theoretically predicted motions to those we actually measure for populations
of {\em Daphnia}. In this case, the individuals are not identical. One way
this population variability could be modeled is by building in a distribution
of power conversion rates as given by the last term on the right in
Eq.~(\ref{eq:2}). In our populations, some Daphnia are young and presumably
cannot convert energy at the same rate and thus cannot swim as fast as older
individuals. We thus modify Eq.~(\ref{eq:3}) by allowing the strength of the
power dissipated in motion to become noisy,
\begin{equation}
  \label{eq:6}
  d\rightarrow d_0+\eta(t)\,,
\end{equation}
where $\eta(t)$ is noise, which, in contrast to the external fluctuation,
$\vec{\xi}(t)$, is an internal fluctuating variable within the population,
meaning that different individuals have different values of $d$. We call this
the population noise. It is Gaussian distributed with zero mean, has standard
deviation $\sigma_d$, and is delta correlated (see also \citet{ErEbSchiSchw99},
where a factor describing the efficiency of conversion is added to the model).
This translates into a noisy dissipation as given by Eq.~(\ref{eq:3}). Using
Eqs.~(\ref{eq:1})and (\ref{eq:3}) with (\ref{eq:6}), we can obtain
trajectories and measures directly comparable to the experimentally measured
ones as we show in the next section below. Here we calculate the probability
densities of velocity for two conditions of noise as shown in
Fig.~\ref{fig:3}. The solid line represents density with the noise in
Eq.~(\ref{eq:1}) only (the external noise) and the dashed line represents both
noises accounting for external and population variability.
The net result of including population variability in the theory is that there
is now a distribution of both radii and velocity at which the particles
travel, a quantity that can be directly measured in the experiment. Note that
for some of the creatures the energy conversion is not efficient enough to
reach the limit cycle. These participate in Brownian motion around the
fixpoint (see inset of Fig.~\ref{fig:2a}). This is indicated by the maximum
around zero in both graphs of Fig.~\ref{fig:3}.

\section{About {\em Daphnia}}\label{sec:about-daphnia}

\begin{figure}[htbp]
  \begin{center}
     \epsfig{file=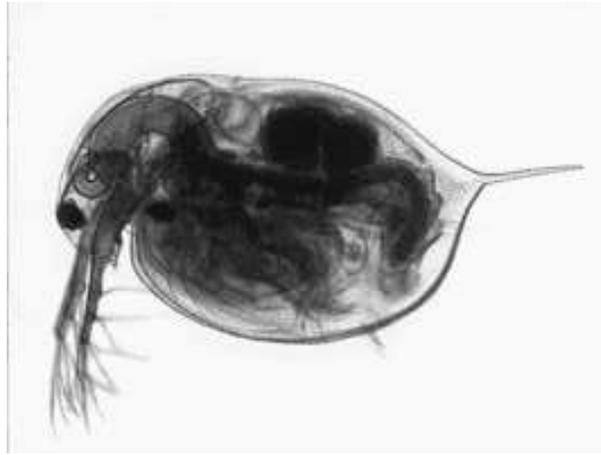,width=6cm,angle=90}
    \caption{A single {\em Daphnia magna} with a store of eggs under the 
      carapace. The two large antennae enable the animal to swim. An array of
      smaller appendages enables feeding. The animals range in size from about
      3 to 5 mm in length.}
    \label{fig:4}
  \end{center}
\end{figure}

{\em Daphnia} are found in many species in fresh water over the entire globe.
Figure~\ref{fig:4} shows a single individual.
They have a long evolutionary history and consequently have developed a range
of complex behaviors that insure their survival in various environments both
stressed and nonstressed. {\em Daphnia} are known to heavily depend on
phototaxis, for example when searching for food \citep{YoGe87}, to confuse and
avoid predators \citep{JeJaKl98} or for performing diel vertical migration
\citep{ZaSu76,RhPaTo01}, a behavior whereby {\em Daphnia} swim near the water
surface while feeding at night, but return to the depths during the day in
order to avoid visual predators (most planktivore fish). They are strongly
attracted to light in the visible range (VIS), repelled by ultraviolet (UV)
and blind to infrared (IR), three facts that we will make use of in the
experimental design outlined below.

\begin{figure}[htbp]
  \begin{center}
    \epsfig{file=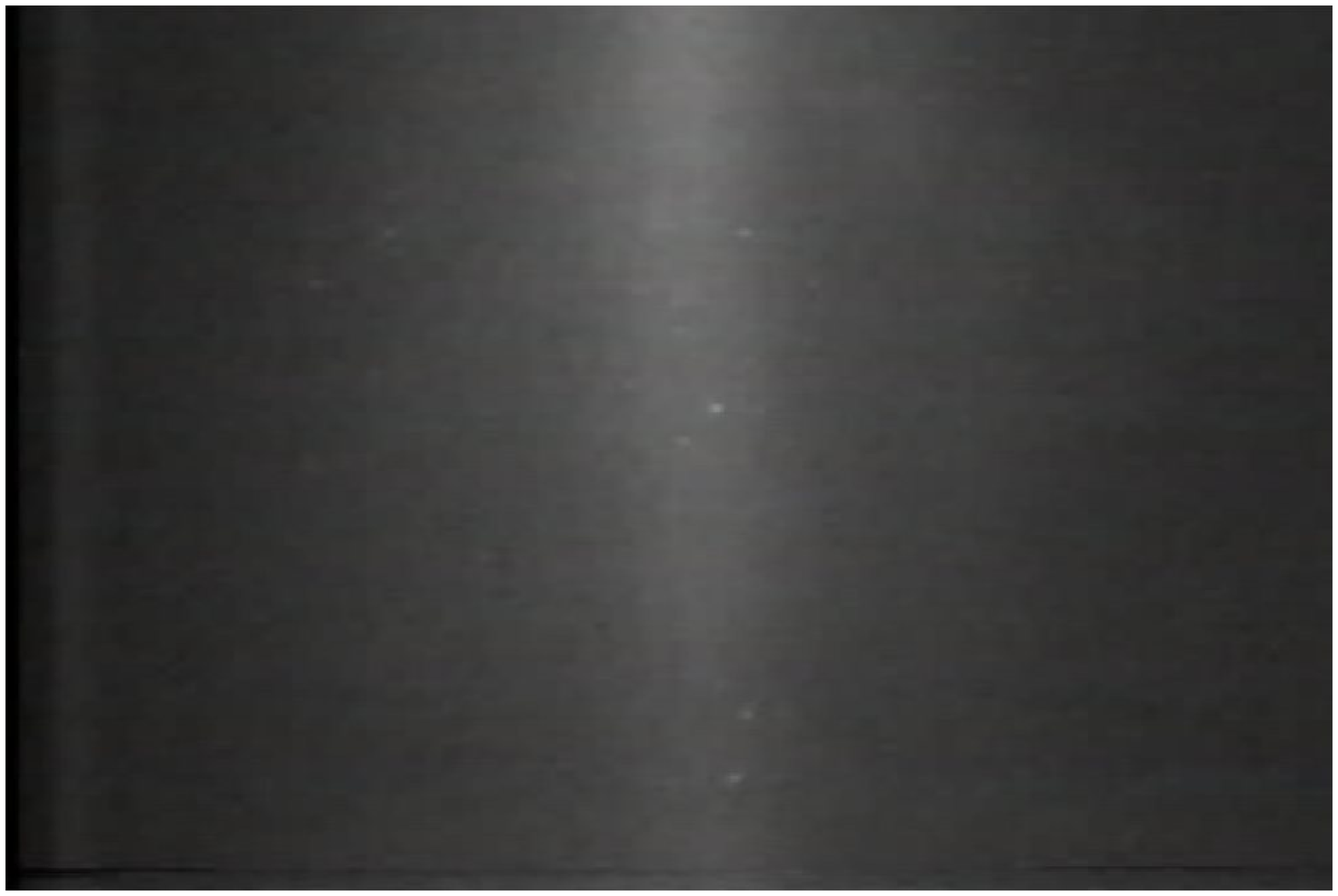,width=4.5cm}
    \epsfig{file=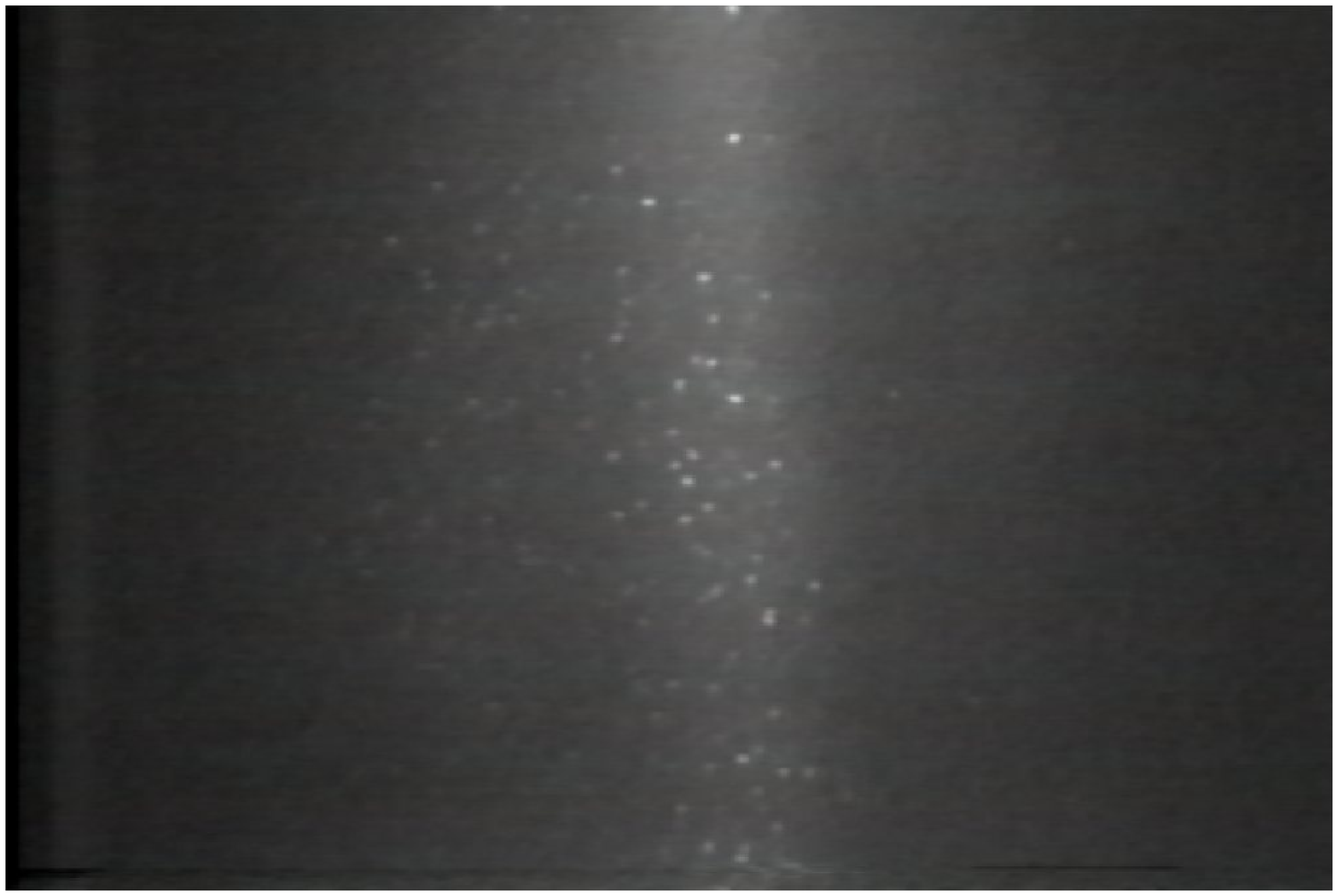,width=4.5cm}
    \epsfig{file=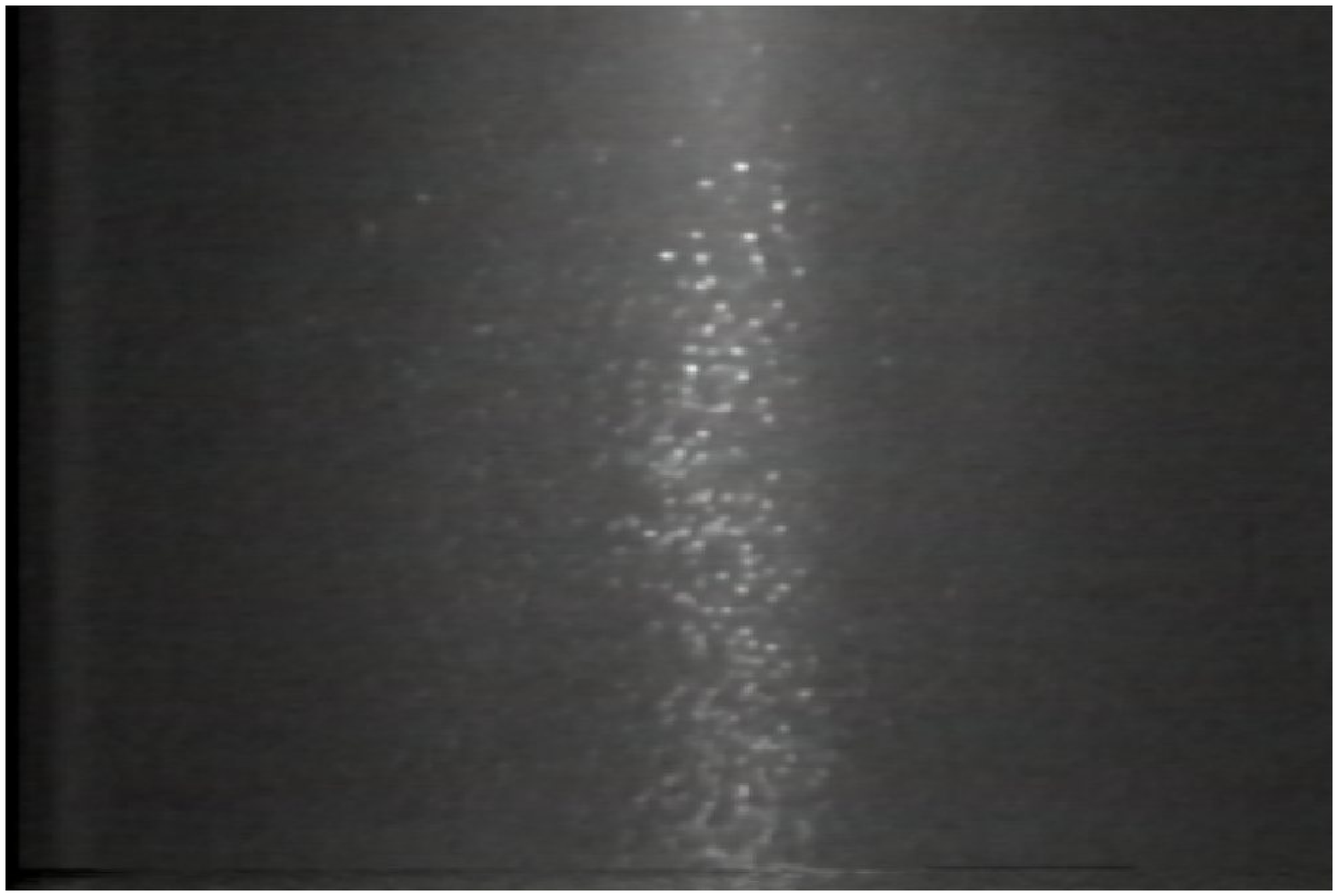,width=4.5cm}
    \caption{{\em Daphnia} are attracted to VIS light as shown by this 
      vertical flashlight beam. Three side view snapshots taken at 20 second
      intervals show the gathering of {\em Daphnia} (white dots) in the shaft
      of light.}
    \label{fig:5}
  \end{center}
\end{figure}

\begin{figure}[htbp]
  \begin{center}
    \epsfig{file=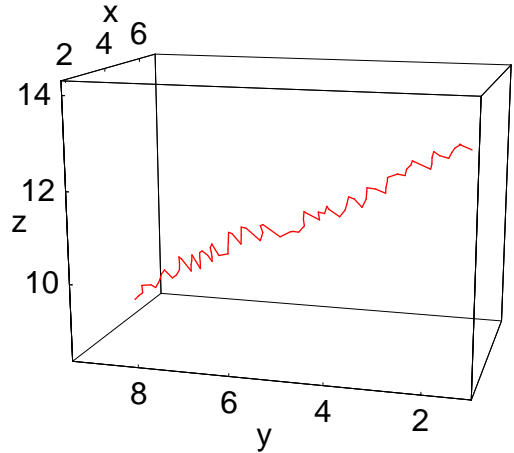,width=8cm}
    \caption{Typical trajectory measured on a {\em Daphnia} in darkness in 
      three dimensions. This trajectory illustrates the hop and sink
      motion with a change of direction between successive hops.}
    \label{fig:6}
  \end{center}
\end{figure}

It is not likely that {\em Daphnia} can form an image with their eyes
according to \citet{BuGo81}. Unlike birds and fish, no direct visual alignment
between {\em Daphnia} has been detected \citep{OkLe02}. They can, however,
determine wavelength, intensity, and direction of light \citep{SmMa90}. {\em
  Daphnia} also use chemotaxis \citep{LaDo93} and can detect water motions
with their mechanoreceptors \citep{HaYa95}. Because there is no known
mechanism of long-range communication between {\em Daphnia} \citep{LaDo93} in
order to display coherent motions they must be individually attracted to a
landmark such as a shaft of VIS light \citep{JeJaKl98}. This can be easily
demonstrated with a vertical flashlight beam in a dark aquarium as shown in
Fig.~\ref{fig:5}.
The swimming behavior of {\em Daphnia} is dominated by the fact that they live
in a relatively low Reynolds number environment \citep{Za80}. They move with
an average `hopping' rate of approximately three moves per second and their
overall speed is 4-16 mm/s with a sinking rate of approximately 3 mm/s
\citep{Do96}. See an example 3-dimensional trajectory in Fig.~\ref{fig:6}.
{\em Daphnia} at low density (about one individual per liter), swimming in the
dark, do not show evidence of any long-range interactions. (There is, however,
a short-range repulsion leading to avoidance maneuvers during close
encounters.) Therefore, their motion can be modeled by the ABP theory of
single self-propelled particles moving in a central field and subject to
noise.

\section{Experiments with small {\em Daphnia} density}
\label{sec:exper-with-small}

\begin{figure}[htbp]
  \begin{center}
    \epsfig{file=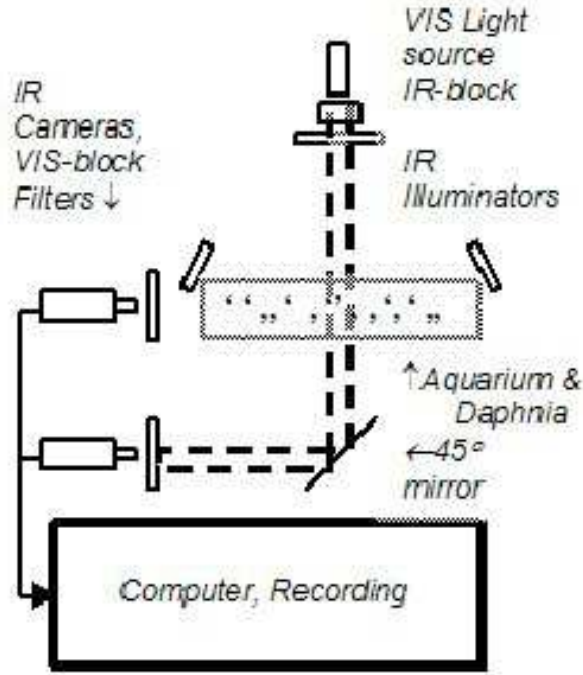,width=8cm}
    \caption{The apparatus, showing the light source and cylindrical VIS light
      shaft, the IR illuminators and cameras. {\em Daphnia} motions can be
      observed and recorded simultaneously in side and bottom views.
      Appropriate filters are arranged such that {\em Daphnia} are recorded in
      IR while responding to and seeing only VIS.}
    \label{fig:7}
  \end{center}
\end{figure}

The apparatus designed to investigate the aforementioned motions is shown in
Fig.~\ref{fig:7}.
It consists of a rectangular aquarium (50~$\times$~20cm, water level 25cm)
with a vertical translucent plastic tube (30 cm long by 1 cm diameter) mounted
in the center. The tube is illuminated with VIS light from above by a fiber
optic cable thus forming a shaft of light propagating radially outward. This
forms a vertical marker with cylindrical symmetry to which the {\em Daphnia}
are attracted. A filter to remove IR from the light shaft is placed between
the illuminator and the fiber optic cable. {\em Daphnia} within the aquarium
are illuminated by IR from arrays of filtered IR-Light Emitting Diodes located
on either side as shown. IR sensitive video cameras (Baxall, Cohu) fronted by
filters to remove VIS light obtain side and bottom views of the aquarium. The
bottom view is afforded by a 45{\tc\symbol{176}} first surface mirror. Images
from the cameras are combined on a split screen viewer, recorded
simultaneously on magnetic tapes, then later digitized and saved on large
capacity disks in a PC. The trajectories are analyzed with the tracking
software (Chromotrack{\tc \symbol{174}} Version 4.02 from San Diego
Instruments, or TrackIt from IguanaGurus). In a later improvement, the video
cameras were replaced by an IR capable camcorder (Sony DCRTRV80) with filters
to block VIS. With this arrangement digital data can be streamed directly to
the computer.

\begin{figure}[htbp]
  \begin{center}
    \subfigure[]{\label{fig:8a}
      \epsfig{file=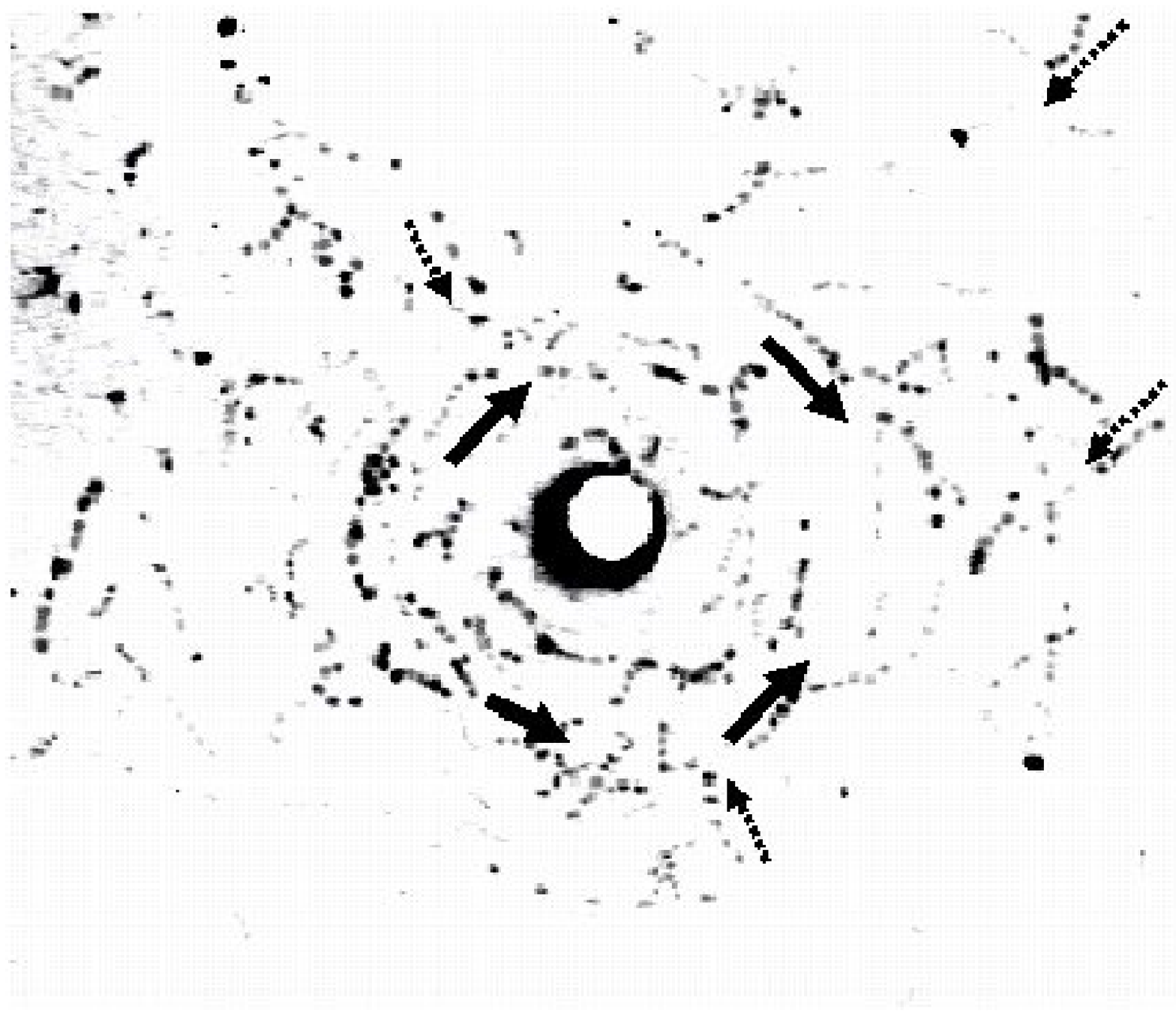,height=4.5cm}}
    \subfigure[]{\label{fig:8b}
      \epsfig{file=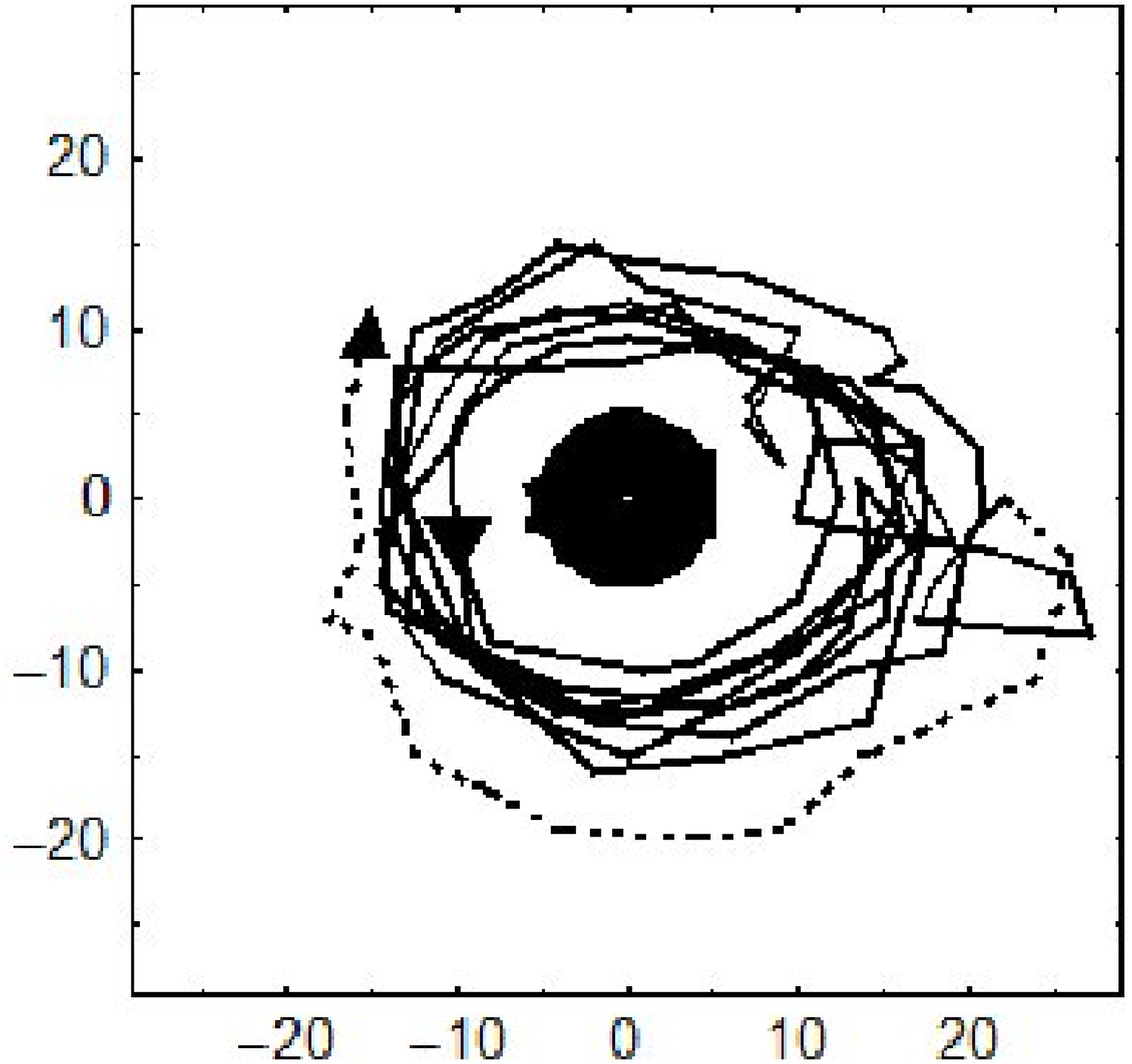,width=4.5cm}}
    \caption{Bottom view of individual {\em Daphnia} motions around the light
      shaft.  (a) About 40 distinguishable individuals, visible as a chain of
      five black dots for each animal, (grey scale inverted for easier
      recognition, time interval between single dots 0.3 sec) are engaged in
      circular motions indicated by filled arrows with approximately equal
      numbers traveling CCW and CW. The 1 cm diam. white circle is the end
      view of the light shaft with VIS light blocked. The dashed arrows in the
      outer areas of the field of view indicate {\em Daphnia} that are moving
      towards the light.  (b) The track of a single individual making seven
      CCW turns (solid line, start marked by arrow) around the light shaft and
      showing a reversal toward the end of the track (dashed line, end marked
      by arrow, coordinates in mm).}
    \label{fig:8}
  \end{center}
\end{figure}

After placing 20 to 80 {\em Daphnia} in the water with algae ({\em Scenedesmus
  quadricula}) for food and leaving them to acclimate and to distribute
uniformly for at least 15min, the VIS light shaft is switched on.  Being
individually attracted to the optical marker, a substantial number can be
observed to move on circular paths around the light shaft in both directions
(CW and CCW) with frequent reversals of direction. The radii of their tracks
are large enough (typically 1 to 5 cm) to exclude the possibility that this
behavior occurs simply due to the hydrodynamic sensing of an artificial object
presumably with their mechanoreceptors. Experiments show that {\em Daphnia}
can sense static objects at distances not exceeding about one to two body
lengths \citep{HaYa95}. Similar circling behavior was observed in the absence
of any solid object in more recent experiments using a high intensity
flashlight shining vertically into the water instead of the solid light tube
\citep{NiOrLoStr03}. Typical tracks are shown in Fig.~\ref{fig:8a} for about
40 individuals and Fig.~\ref{fig:8b} for a single individual that makes seven
complete CCW turns and a reversal of direction around the central marker.
Arrows indicate the rotational directions. The trajectories look as in the
theory shown in the inset of Fig.~\ref{fig:2b}.
While these results from experiment and theory seem quite similar, statistical
measures are necessary for quantitative comparisons.

An essential point is that single {\em Daphnia} make rotational motions
around the central attractant. This indicates that, for {\em Daphnia} at
least, circling is not an emergent or self-organized motion that occurs as an
inherent property of a swarm of animals \citep{PaEd99}. Circular motions in
zooplankton at low density were observed previously \citep{YoGe87,YoTa90} but
only incidentally in the course of different experiments. Below we show that
high-density swarms of {\em Daphnia} are capable of vortex motions similar to
those observed in bacteria \citep{CziBeCoVi96}, slime molds \citep{RaNiSaLe99}
and fish \citep{PaViGr02}. But in {\em Daphnia} the onset of the vortex is
the result of a symmetry-breaking process that transforms the two symmetric
limit cycle motions (CCW and CW), observable for single individuals in both
theory and experiment, into motion in one direction only. Certainly individual
birds in some cases follow rotational paths around central attractive markers
\citep{LaFr88}. Possibly bacteria and fish do also. It is therefore not
demonstrated that the vortex motion sometimes observed in swarms of these
animals is an ``emergent'' property unrelated to the motions of individuals.

\begin{figure}[htbp]
  \begin{center}
    \subfigure[]{
      \epsfig{file=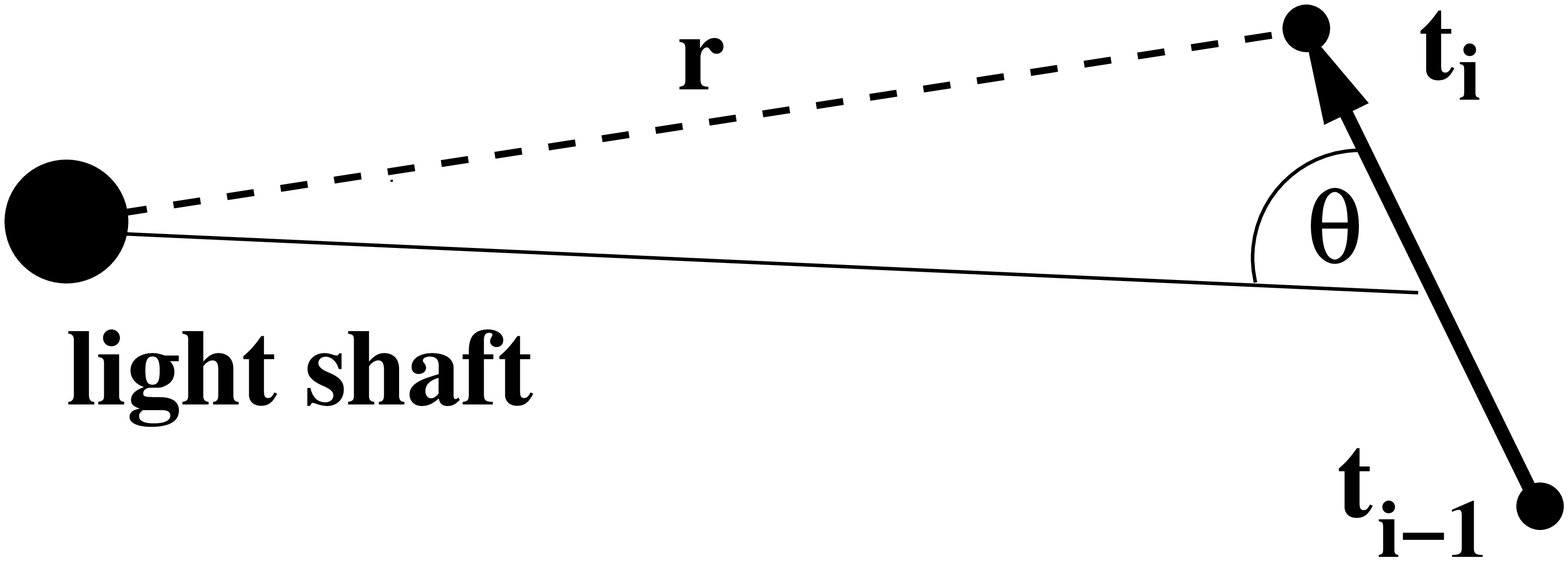,width=5cm}
      \label{fig:9a}}
    \subfigure[]{
      \unitlength 0.3mm
      \def\epsfsize#1#2{0.36#1}
      \begin{picture}(250,150)
        \put(0,15){\epsfbox{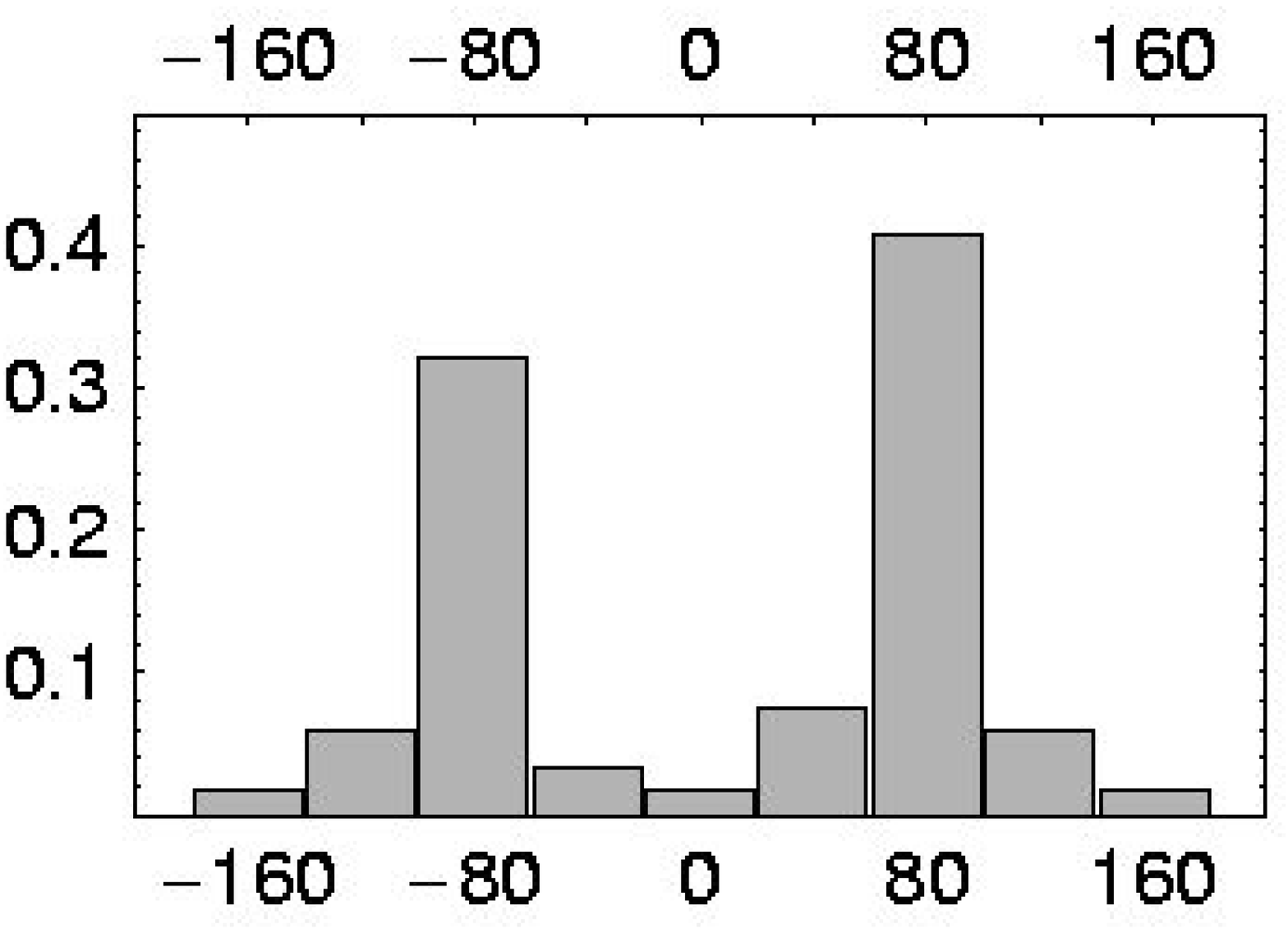}}
        \put(23,148){\epsfig{file=fig/09/whitespace.eps,height=5mm,width=70mm}}
        \put(-2,78){\makebox(10,10)[r]{\rotate[l]{\fontsize{15}{15}
              \selectfont{$\displaystyle P(\theta)$}}}} 
        \put(128,0){\makebox(10,10)[r]{{{\fontsize{15}{15}
                \selectfont{$\displaystyle  \theta $}}}} }
      \end{picture} 
      \label{fig:9b}}
    \caption{The heading angle. (a) Definition of the heading angle, $\theta$,
      as measured for every hop movement of an individual. (b) Histogram of
      heading angles, $P(\theta)$, obtained from 624 hops from the tracks of
      four individuals. Plot (b) is adapted from \citet{OrBaMo02a}.}
    \label{fig:9}
  \end{center}
\end{figure}

We turn now to statistical measures of {\em Daphnia} motions at small
density. The dominant part of the motion of single {\em Daphnia} in the light
field takes place in the horizontal plane. Indeed individuals tend to circle
in layers with little vertical migration, as we have observed in 3-dimensional
tracks obtained in the laboratory of J. R. Strickler \citep{Ni02}.
%% \citep{NiOr02}
It is therefore sufficient to restrict the characterization of the observed
circling behavior to two dimensions, analyzing only the bottom-view pictures
of the aquarium. To quantify the amount of circling, we measured the
probability distribution $P(\theta)$ of the heading angles for several {\em
  Daphnia} observed circling around the light shaft. The heading angle,
$\theta$, is defined as the angle between the vector direction of a single hop
and the direction to the light shaft measured in the middle of each hop as
shown in Fig.~\ref{fig:9a}.
This distribution is double peaked with maxima located at approximately
$\pm$90{\tc\symbol{176}}. It indicates two properties of the motion: First,
the twin peaks at $\pm$90{\tc\symbol{176}} demonstrate that the average motion
is resolved into two rotational directions, CW ($+$90{\tc\symbol{176}}) and
CCW ($-$90{\tc\symbol{176}}). Second, the magnitudes of the two peaks are
nearly equal, demonstrating that the CW ($+$90{\tc\symbol{176}}) and CCW
($-$90{\tc\symbol{176}}) motions occur with approximately equal probabilities.
Both of these experimental observations are predicted by the ABP theory.
        
\begin{figure}[htbp]
  \begin{center}
    \subfigure[]{
      \unitlength 0.25mm
      \def\epsfsize#1#2{0.36#1}
      \begin{picture}(250,150)
        \put(0,15){\epsfbox{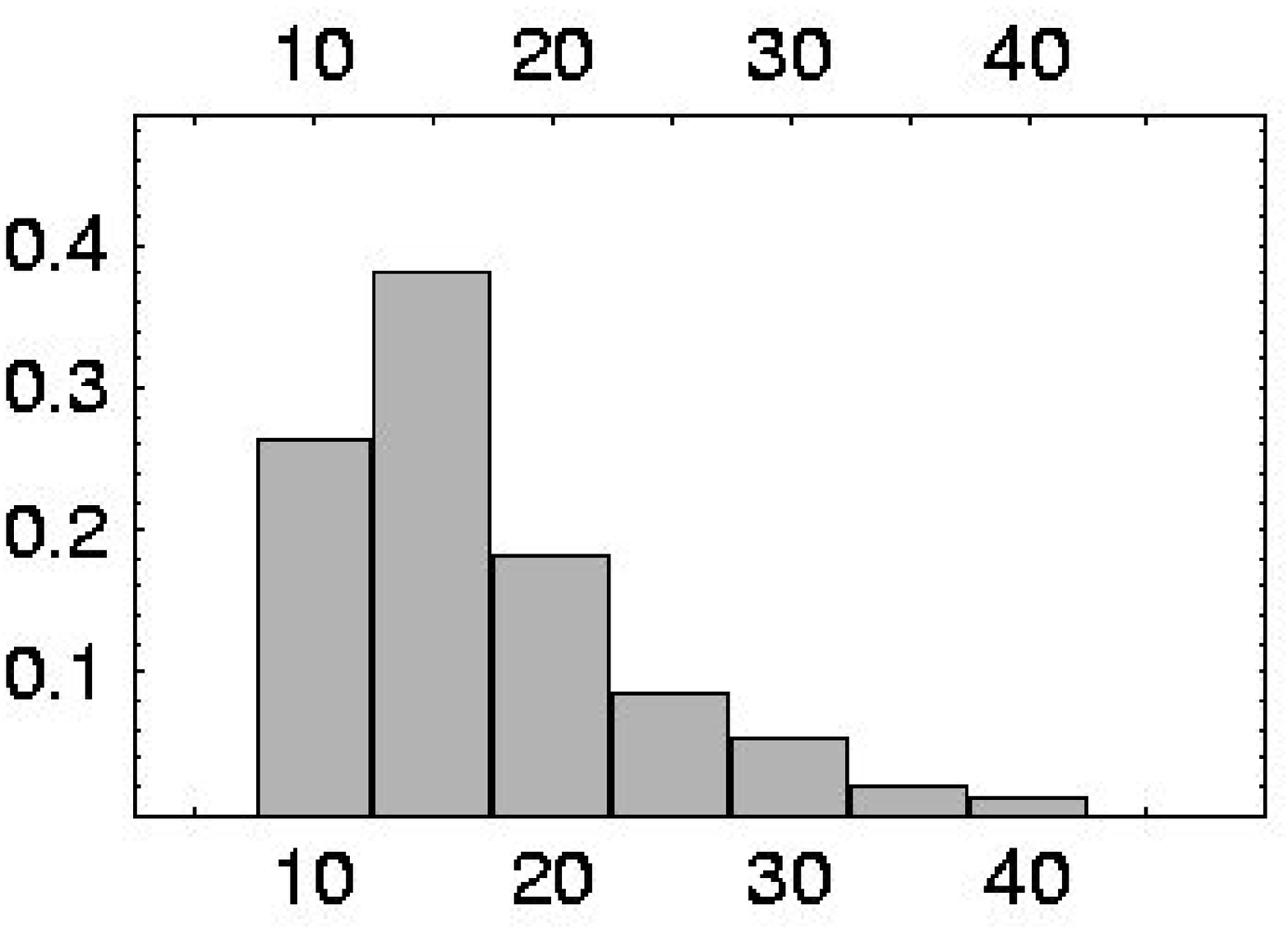}}
        \put(23,175){\epsfig{file=fig/09/whitespace.eps,height=5mm,width=70mm}}
        \put(-2,78){\makebox(10,10)[r]{\rotate[l]{\fontsize{12}{12}
              \selectfont{$\displaystyle P(r)$}}}} 
        \put(128,0){\makebox(10,10)[r]{{{\fontsize{12}{12}
                \selectfont{$\displaystyle  r $}}}} }
      \end{picture} 
      \label{fig:10a}}
    \hfill
    \subfigure[]{
      \unitlength 0.25mm
      \def\epsfsize#1#2{0.36#1}
      \begin{picture}(250,150)
        \put(0,15){\epsfbox{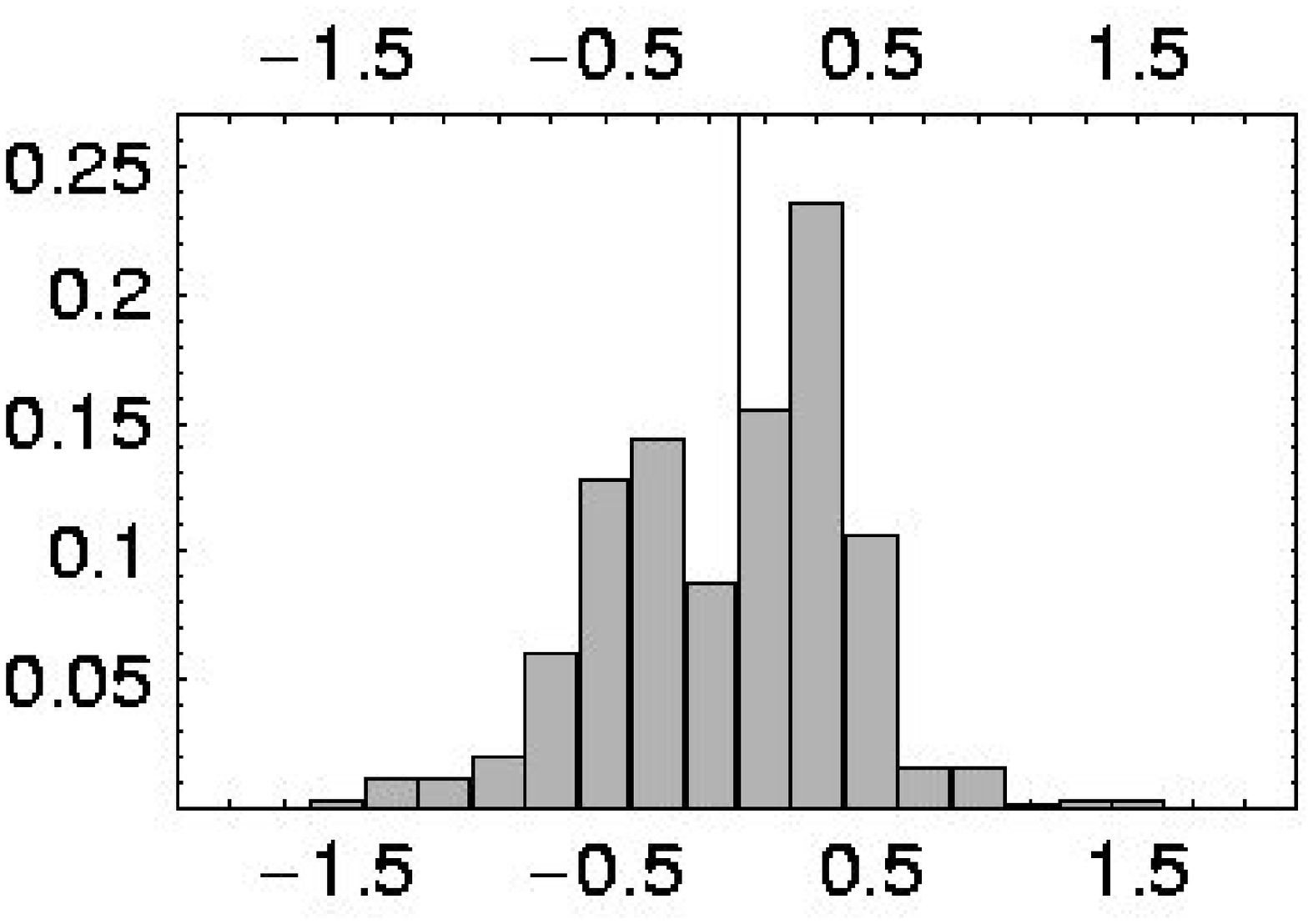}}
        \put(23,175){\epsfig{file=fig/09/whitespace.eps,height=5mm,width=70mm}}
        \put(-2,78){\makebox(10,10)[r]{\rotate[l]{\fontsize{12}{12}
              \selectfont{$\displaystyle P(L_{\mathrm{ang}})$}}}} 
        \put(137,0){\makebox(10,10)[r]{{{\fontsize{12}{12}
                \selectfont{$\displaystyle  L_{\mathrm{ang}}$}}}} }
      \end{picture} 
      \label{fig:10b}}\\[0.5cm]
    \subfigure[]{
      \unitlength 0.25mm
      \def\epsfsize#1#2{0.36#1}
      \begin{picture}(250,150)
        \put(0,15){\epsfbox{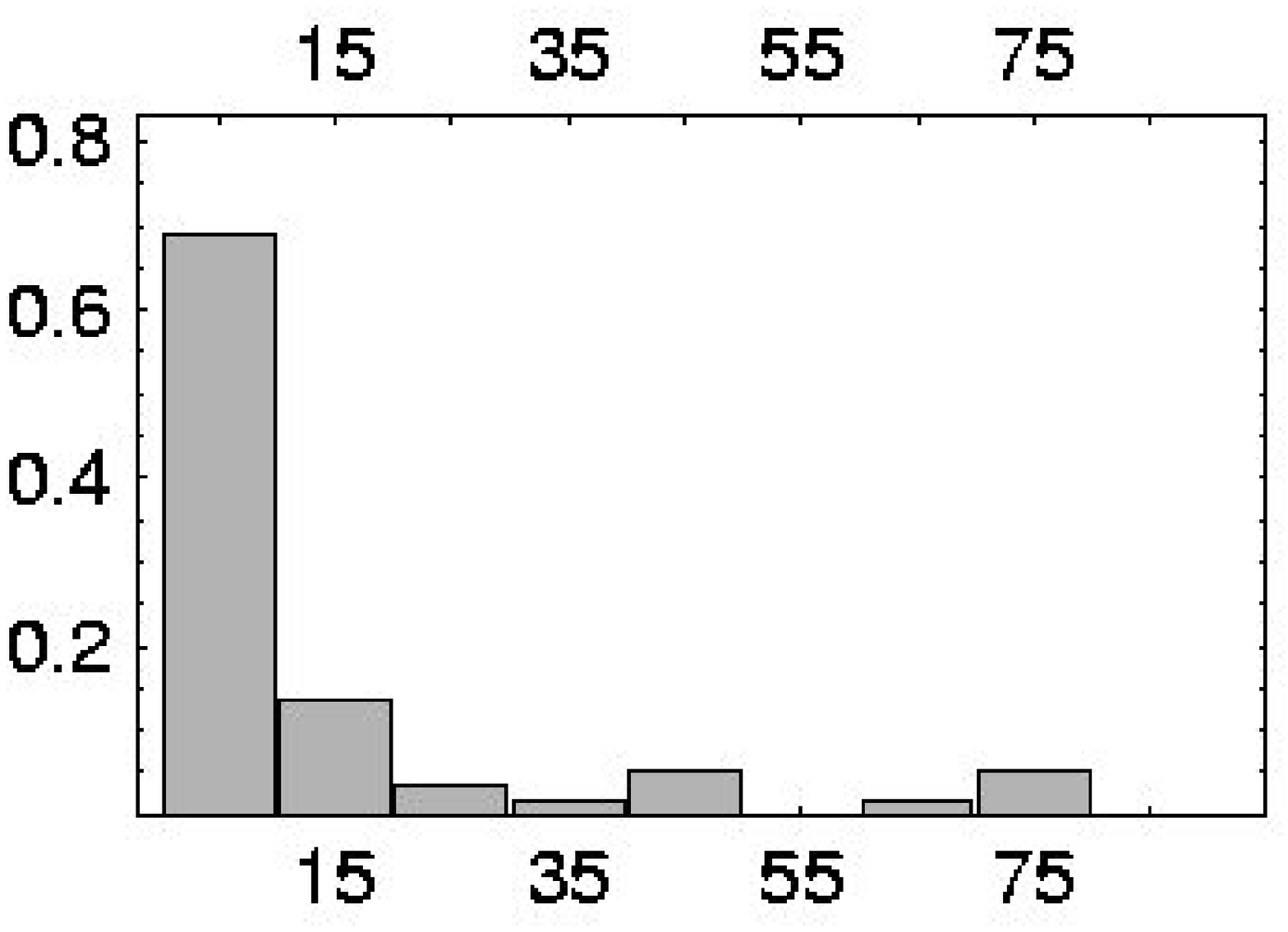}}
        \put(23,175){\epsfig{file=fig/09/whitespace.eps,height=5mm,width=70mm}}
        \put(-2,78){\makebox(10,10)[r]{\rotate[l]{\fontsize{12}{12}
              \selectfont{$\displaystyle P(M_{C})$}}}} 
        \put(135,0){\makebox(10,10)[r]{{{\fontsize{12}{12}
                \selectfont{$\displaystyle M_{C}$}}}} }
      \end{picture} 
\label{fig:10c}}
    \caption{Histograms of (a) the circling radii, $P(r)$, (b) the angular
      momentum, $P(L_{\rm ang})$, and (c) the directional reversals, $P(M_{\rm
        C})$. These data were obtained from the same set of hops using the
      same track data as for Fig.~\ref{fig:9}. Plots are adapted from
      \citet{OrBaMo02a,OrBaMo02b}.}
    \label{fig:10}
  \end{center}
\end{figure}

For the same set of hops we obtained a histogram representing the probability
distribution $P(r)$ of the distance $r$ of the {\em Daphnia} to the light
shaft as shown in Fig.~\ref{fig:10a}. Figure~\ref{fig:10b} shows a histogram
of the angular momentum, again with twin peaks of approximately equal
magnitudes. In Fig.~\ref{fig:10c} we show a histogram of the average length of
circling in one direction (CW or CCW) before reversing the direction. Further
on we determined the average length of circling in one direction before
reversing to be 11.8 moves.

\section{The Many Particle ABP Theory}\label{sec:many-particle-abp}

When there are many particles with a long range attractive interaction between
them, the global effect is that the particle-particle interactions can be
replaced by a mean field. This field is the result of a central potential that
causes confinement of the particles. This phenomenon can actually be
demonstrated in experiments with the zooplankton {\em Chydoridae}. A long
range attractive interaction can be switched on by causing one or a very few
individuals to become visible to all the others by shining light on them. When
that happens all animals are attracted to the visible few. The light scattered
from the few must, however, dominate the visual environment. This demonstration
has been realized in the laboratory of J. R. Strickler by directing a narrow,
blue, VIS, laser beam vertically into an aquarium which, apart from the beam,
is dark. The beam itself cannot be seen by the animals in the tank since the
medium is clear and the beam is very narrow. Many animals are initially
distributed either at the bottom or throughout the aquarium. One or a few
individuals by accident swim into the beam, become brightly illuminated, and
thus become visible to all the rest. The remaining individuals are attracted
by this light, immediately begin migrating toward the beam and become
illuminated themselves, thereby intensifying the attraction, a positive
feedback effect whose strength is dependent on the number illuminated (see
Fig.~11 in \citet{Str98} and the movie at
{\texttt{http://www.uwm.edu/\~{}jrs/out\_of\_corners.htm}}). Thus the motion
mimics that of a large number of animals in a central attractive potential,
but in this case arises solely due to a long range, indirect attractive
interaction between the animals. Zooplankton in the dark, or in the
aforementioned aquarium without a single individual in the beam to scatter
light, show no evidence of a direct long range interaction.

In the theory we can add the attractive interaction summed over all particles
and this will enter an equation of motion similar to Eq.~(\ref{eq:1}) but now
representing the balance of forces on one of the particles, $i$, in the
ensemble,
\begin{equation}
  \label{eq:7}
  m\partial_t \vec{v}_i=-\gamma_0\,\vec{v}_i + d\,e_i(t)\,\vec{v}_i
                        -\kappa_{\rm h}\left[\vec{r}_i-\frac 1N \sum_j 
                          \vec{r}_j\right]
                        +\sqrt{2D}\vec{\xi}_i(t)\,,
\end{equation}
and the energy uptake of the $i$-th particle is,
\begin{equation}
  \label{eq:8}
  \partial_t e_i(t)=q_0-c e_i(t) -d\, v_i^2\, e_i(t)\,.
\end{equation}
Equations~(\ref{eq:3}) and (\ref{eq:4}) remain the same except that the
velocity must be indexed: Moreover, if we build in the effect of population
variability, then the parameter, $d \rightarrow d_i$ in
Eqs.~(\ref{eq:3})--~(\ref{eq:5}), where the population noise in
Eq.~(\ref{eq:6}) results in a Gaussian distribution of the $d_i$. Notice that
in Eq.~(\ref{eq:7}) the term in square brackets has replaced the force
generated by external potential $-\nabla U(\vec{r})$. Moreover, the force is
linear in the distance between particles, $\vec{r}_i-\vec{r}_j$, so that the
mean field potential is quadratic as before, and its strength is adjusted with
the coupling constant $\kappa_{\rm h}$.

\begin{figure}[htbp]
  \begin{center}
    \subfigure[]{\label{fig:11a}
      \epsfig{file=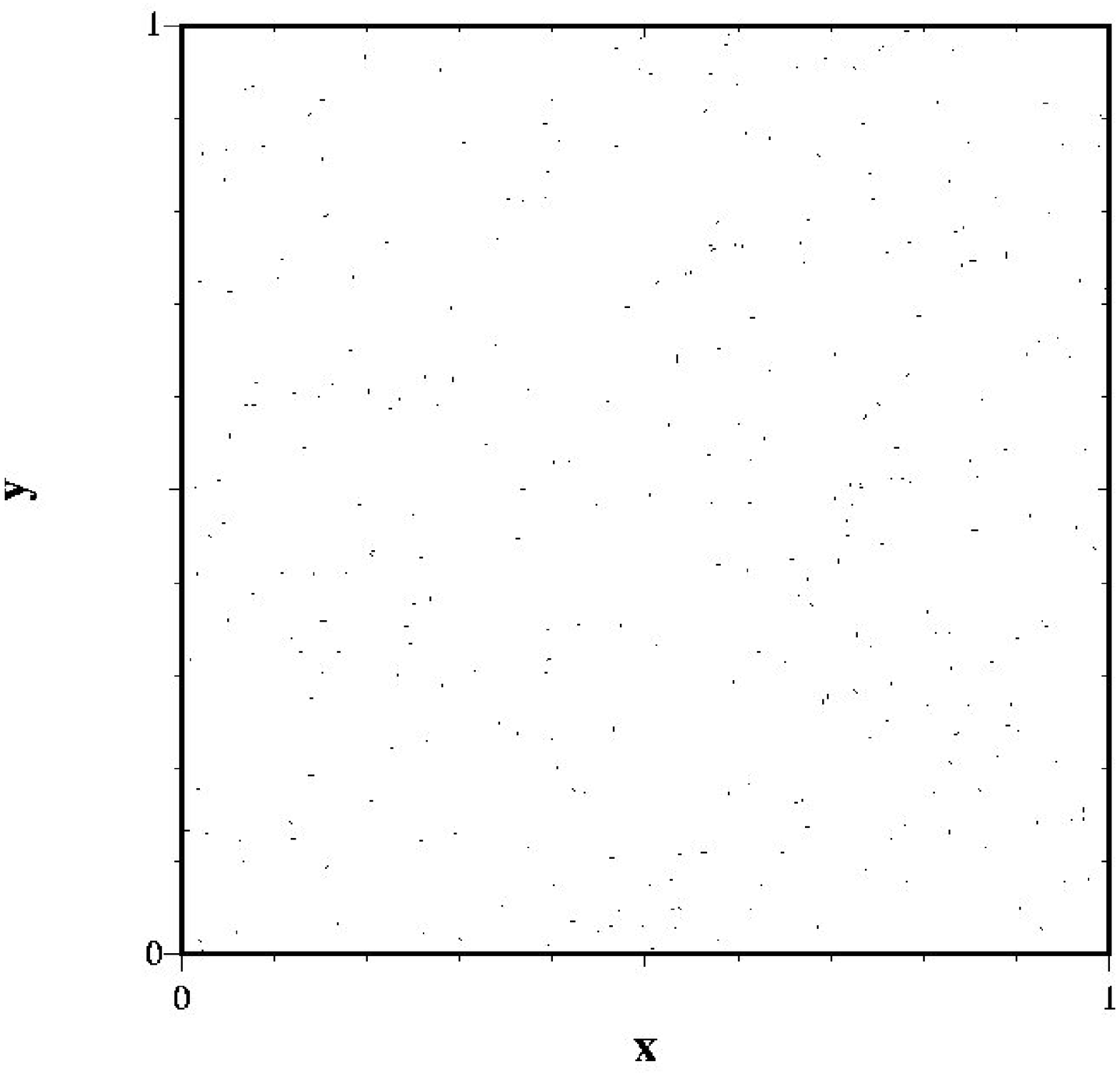,width=3.6cm}}
    \subfigure[]{\label{fig:11b}
      \epsfig{file=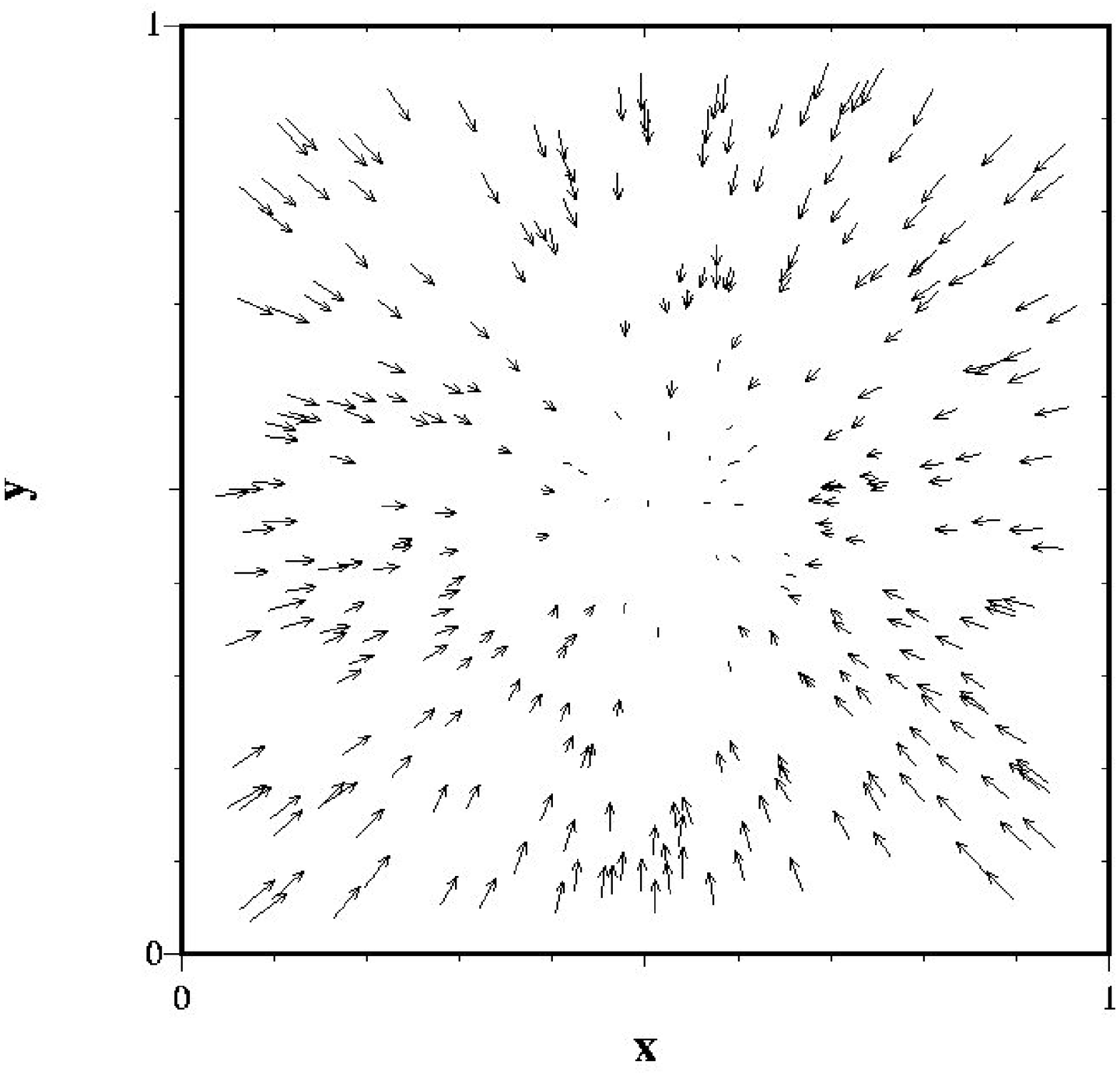,width=3.6cm}}\\
    \subfigure[]{\label{fig:11c}
      \epsfig{file=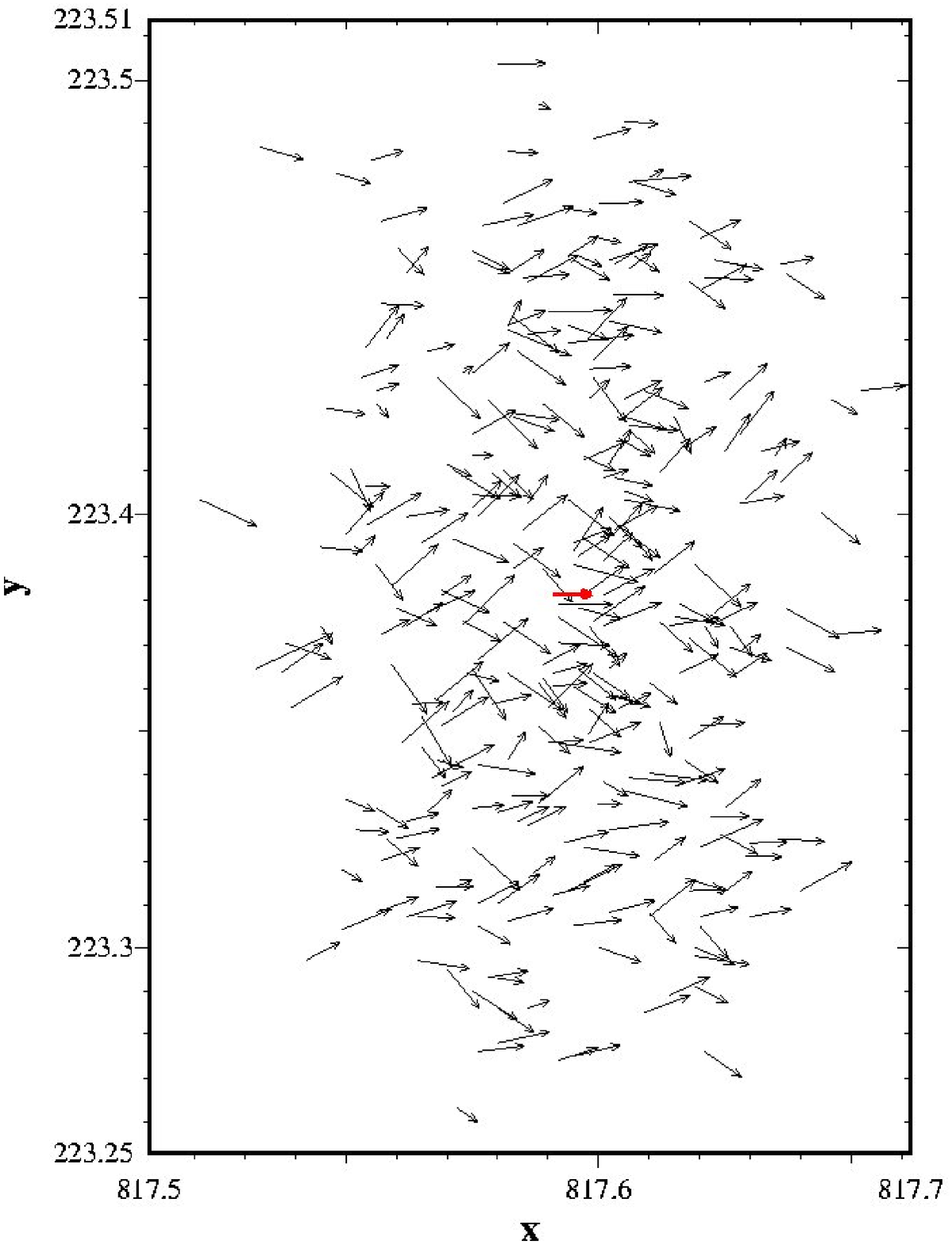,width=3.6cm,height=4.1cm}}
    \subfigure[]{\label{fig:11d}
      \epsfig{file=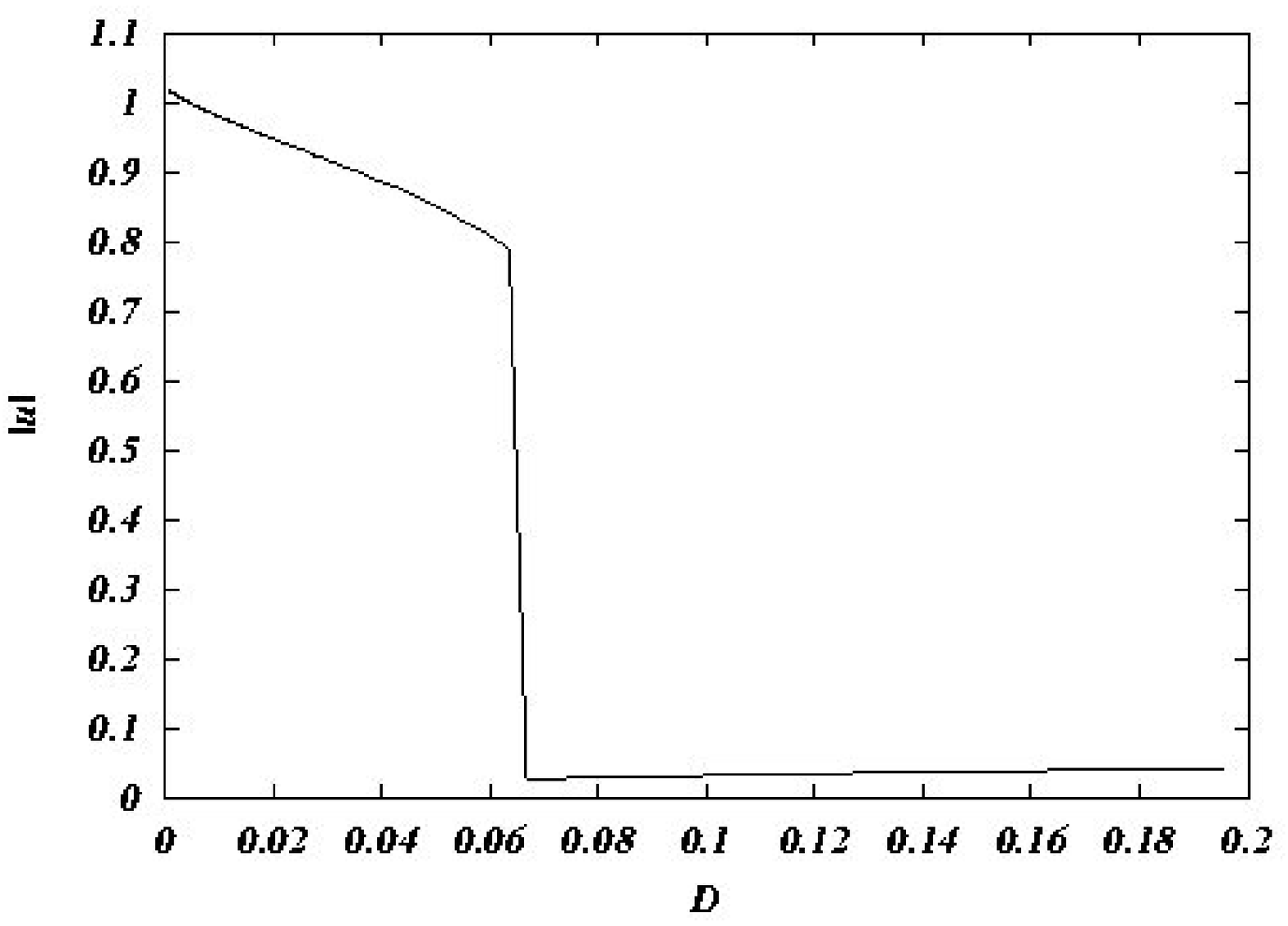,width=6cm}}
    \subfigure[]{\label{fig:11e}
      \epsfig{file=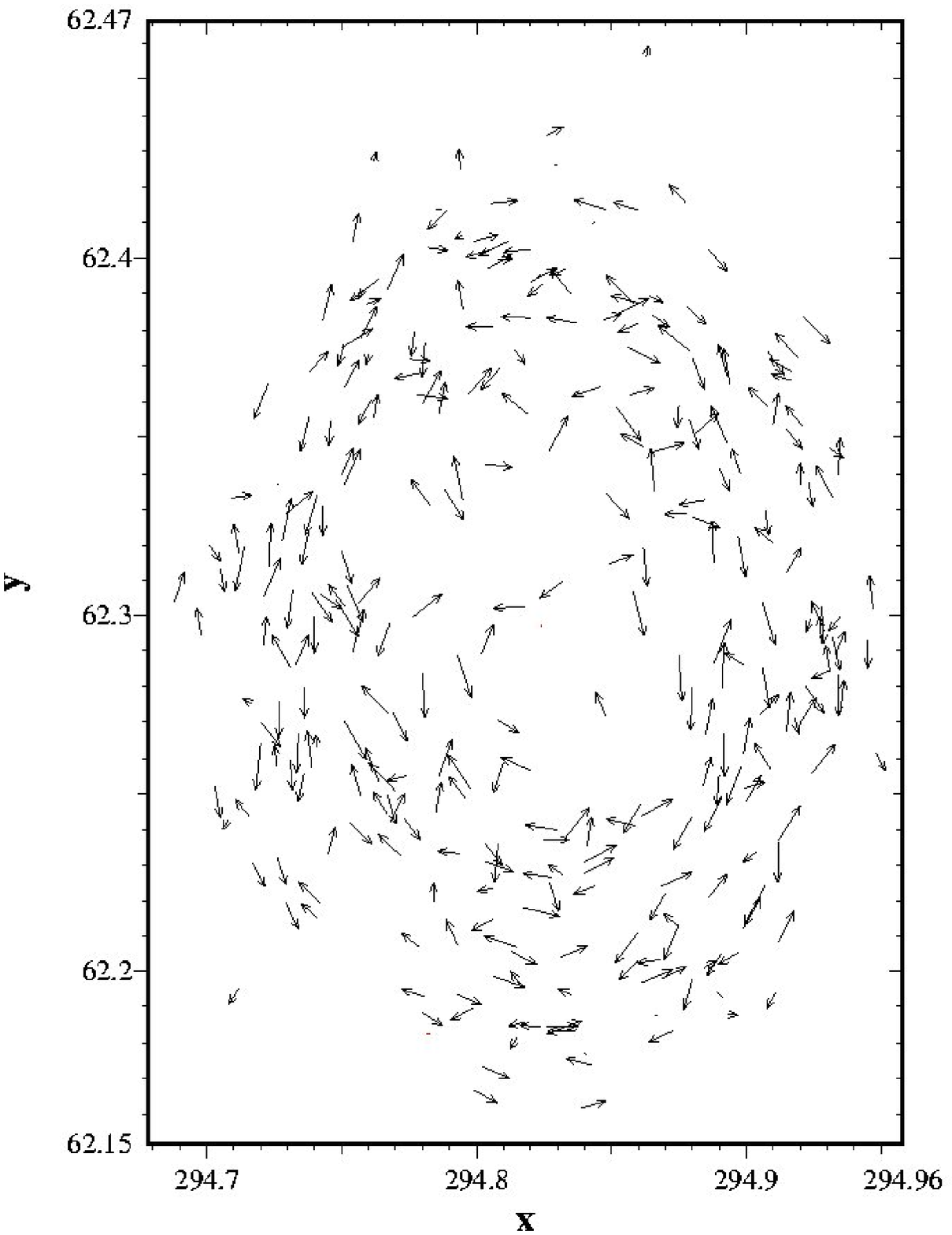,width=3.6cm,height=4.1cm}}
    \caption{Swarming in the many particle ABP theory. (a) Initially many
      particles are widely dispersed to random positions. (b) Under the
      influence of their mean field, they move toward the center of mass. (c)
      Directed motion. (d) Noise induced collapse of translational velocity to
      zero mean translation for $D > D_{\rm crit}$. (e) Both sets of initial
      conditions finally converge on an equilibrium state is established with
      approximately equal numbers of particles rotating CW and CCW. The mean
      radius of the two limit cycle distributions depends on the mean
      self-propelling velocity.}
    \label{fig:11}
  \end{center}
\end{figure}

The many-particle theory as outlined here, leads to results similar to those
predicted by Eqs.~(\ref{eq:3})--~(\ref{eq:5}), in particular, there is a
bifurcation from a noisy fixed point motion to the symmetric pair of limit
cycles except now the entire population participates in these motions. For
simplicity, we treat this problem only considering the environmental noise,
$\vec{\xi}(t)$, appearing in Eq.~(\ref{eq:1}), though in the many particle
case, Eq.~(\ref{eq:7}), an independent noise, $\vec{\xi}_i(t)$, applies to
each particle in the population. The appropriate measures are thus probability
densities in both coordinate and velocity space. In addition to these motions,
the theory predicts the third motion of our suite: swarming. A large
population of particles, initially widely and randomly dispersed, as shown in
Fig.~\ref{fig:11a}, begin to migrate toward the center of mass of the colony
when the inter-particle interactions are switched on as shown in
Fig.~\ref{fig:11b}. Their individual motions are largely rotational, but this
is not visible to the eye, since on average equal numbers rotate in opposite
directions. The CW and CCW motions are marked by arrows in the Figure.
Figure~\ref{fig:11} shows example trajectories but the appropriate measures
are probability distributions of spatial and velocity coordinates on the
$x$-$y$ plane.

The simulations of the many particle theory show another interesting phenomenon
that is observable for a different set of initial conditions as shown in
Fig.~\ref{fig:11c}. The particles are still initially randomly distributed
over the plane, but now each is given the self-propelling velocity and all
vectors are initially parallel (in this case pointed in the $x$ direction). We
call this directed motion. For noise intensity smaller than some critical
value, $D < D_{\rm crit}$, the particles continue to move with a mean
translational velocity, but for $D > D_{\rm crit}$, the mean translational
velocity rapidly descends to zero and the motion is all rotational (again with
equal probability for CW and CCW motions). This is a new type of noise induced
transition \citep[see e.g.][]{HoLe84} and it is shown in Fig.~\ref{fig:11d}.
The critical noise intensity, $D_{\rm crit}$, depends on the parameters of the
velocity dependent dissipation, $\gamma(v)$, and has been calculated in the
limit of large population density \citep{ErEbMi01}. For both sets of initial
conditions, the final state is the set of two symmetric counter-rotating limit
cycles, as shown by the example trajectories in Fig.~\ref{fig:11e}.

\section{Experiments with large {\em Daphnia} density}
\label{sec:exper-with-large}

Swarms are often formed for different purposes, such as enhanced feeding,
mating and offspring rearing, by self-propelled animals living and moving in
three-dimensional environments. Besides this, many animals are thought to form
swarms to confuse and avoid predators \citetext{\citealp{HuDr67,Ha71,Pu73},
  see also \citealp{PaEd99} and \citealp{OkLe02}}. Swarming for this purpose
seems common to a number of animals including planktivore fish, for example,
sardines, mackerel and anchovetta \citep{Pa82,HaWaMa86}, zooplankton
\citep{JaBiJo94,KvKl95} and some species of birds \citep{CaMaPu80}. Such
swarms, occasionally also associated with rotational motions, have been called
self-organized or emergent structures \citep[see
e.g.][]{PaEd99,CaDeFrSnThBo01,OkLe02} meaning that there is no single leader
nor a dynamical process that governs the motions of individual animals in such
a way as to give rise to the swarm structure.

We can consider swarming to be simply a local increase in density of animals
without a net global coherent motion. An attractant, such as a light marker is
usually necessary \citep[e.g.][]{JaJo88,BuPeAm96}.  An additional collective
motion, however, has been observed in swarms.  According to anecdotal
evidence, transitions from more-or-less random motions to an average, coherent
rotational motion have sometimes been observed in the field.  These are called
vortex motions and have actually been described in detail for a species of
oceanic plankton \citep{LoRa86}. In the laboratory, vortex motions have been
studied in bacterial colonies \citep{BeShoTeCoCziVi94,CziBeCoVi96} and slime
molds \citep{RaNiSaLe99} but have not been well investigated for the larger
animals largely due to limitations imposed by the size of the swarms, for
example of birds and fish.  The physical, biological, and chemical reasons for
vortex-swarming are, consequently, not well understood. But the theories
discussed here can provide at least a minimal set of conditions necessary for
vortex formation. The symmetry of the pair of limit cycles must be broken in
order for the vortex to form. Symmetry-breaking processes based on pair
avoidance (arising for example from short range repulsive potentials) are
discussed in Sect.~\ref{sec:how-symmetry-broken} below. In
Sect.~\ref{sec:outline-rwt}, we also consider hydrodynamic feedback as a
symmetry breaking mechanism.

\begin{figure}[htbp]
  \begin{center}
    \subfigure[]{\label{fig:12a}
      \epsfig{file=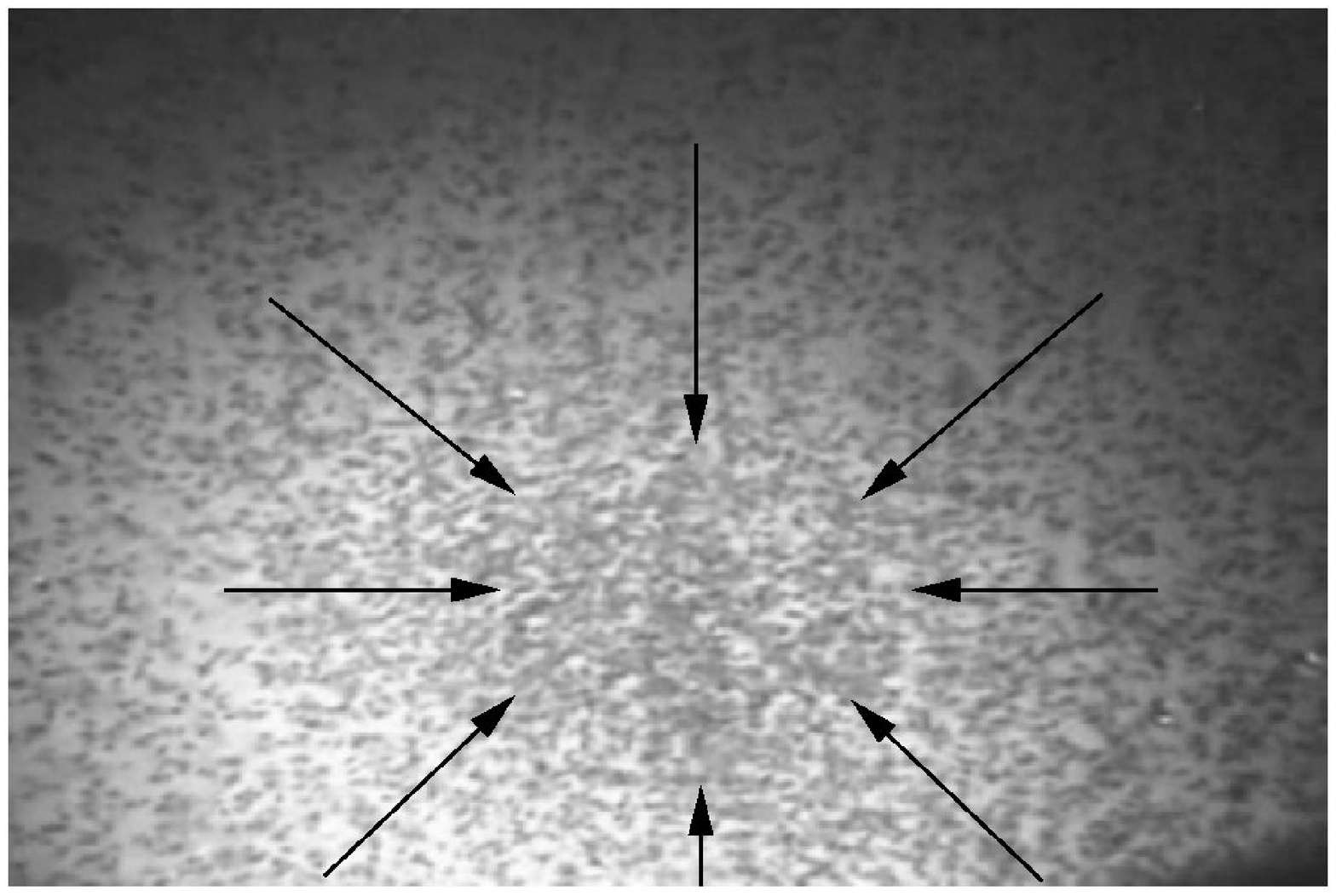,width=4.4cm}}
    \subfigure[]{\label{fig:12b}
      \epsfig{file=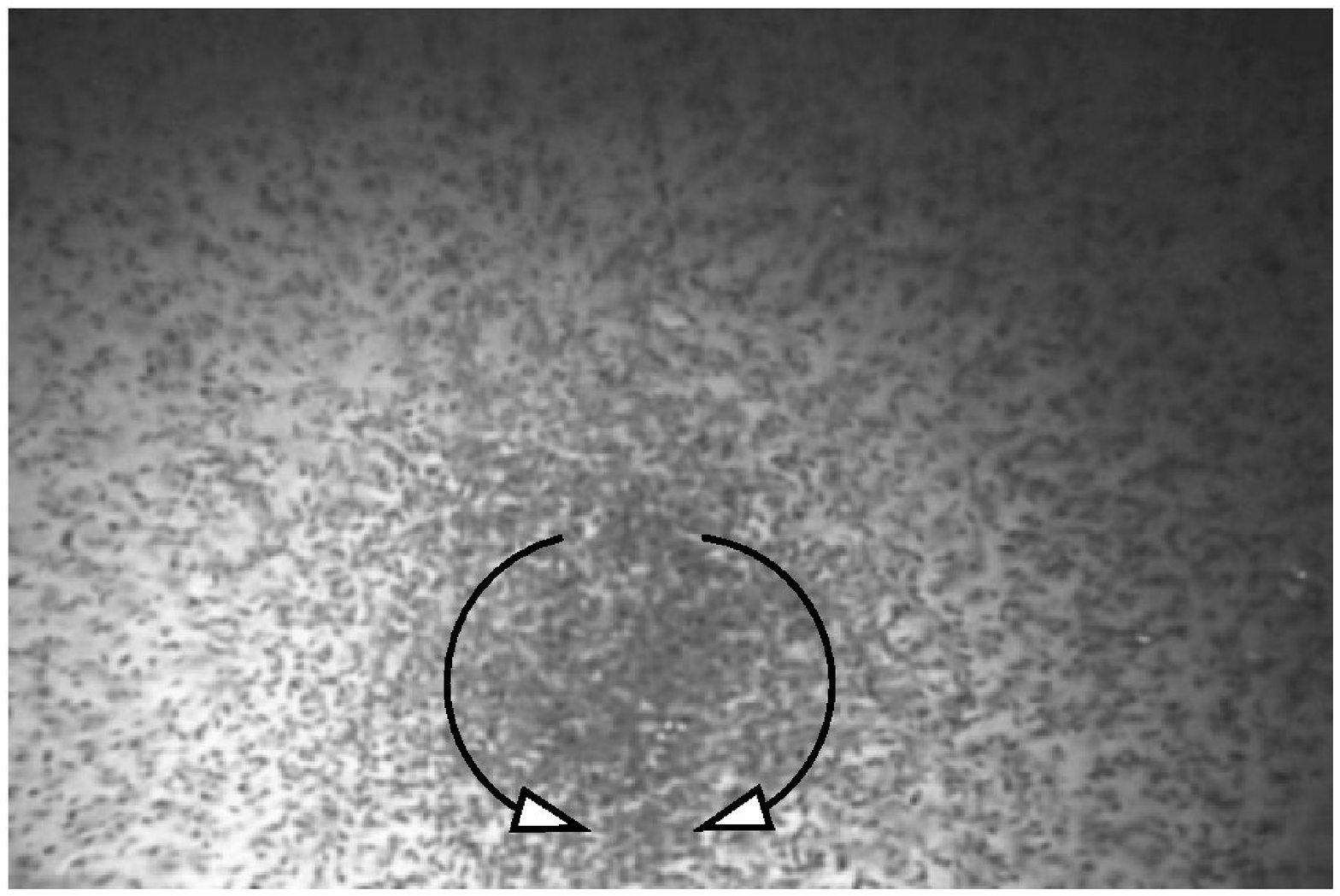,width=4.4cm}}
    \subfigure[]{\label{fig:12c}
      \epsfig{file=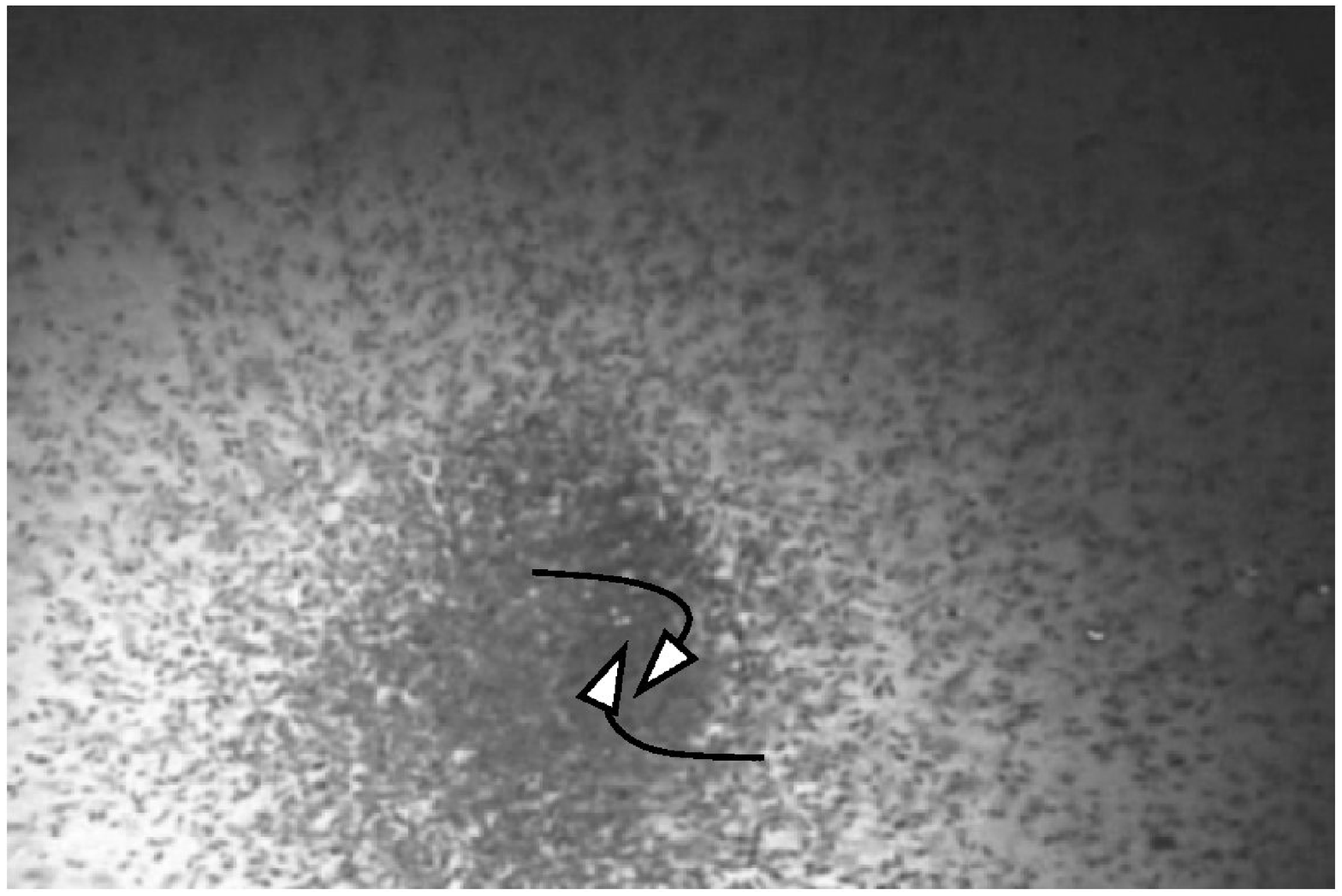,width=4.4cm}}
    \caption{Three top view snapshots taken at 5 second intervals show the 
      formation of a {\em Daphnia} vortex-swarm. (a) {\em Daphnia} gathering
      at the light shaft, taken 5 seconds after switching on the visible
      light, (b) rotational motion of swarming {\em Daphnia} in both
      directions, CCW and CW, and (c) vortex-swarming in CW direction.}
    \label{fig:12}
  \end{center}
\end{figure}

We have observed transitions to vortex motions in {\em Daphnia} swarms using
the apparatus shown in Fig.~\ref{fig:7} as well as in a cylindrical aquarium
(about 50cm diameter) recorded with a digital camera from above. The vertical
visible light shaft, provided either by a flashlight beam or by the cylinder
of Lucite, is the central attractant. First, {\em Daphnia} are conditioned in
the dark for typically 15 minutes in order that they assume an approximately
random distribution. Then the light shaft is switched on. The following
motions are typically observed. During the first few seconds, the animals
swarm toward the light shaft as shown in Fig.~\ref{fig:12a}. During this
period no rotational motions can be observed; the animals simply rush toward
the attractant. In the next period rotational motion becomes evident but with
approximately equal numbers of individuals traveling in both directions (see
Fig.~\ref{fig:12b}).  Finally the individuals tend to align their velocities,
and global rotational motion either in the CW or CCW direction occurs as shown
in Fig.~\ref{fig:12c}.
We have observed that the times for these processes to take place strongly
depend upon the {\em Daphnia} density, being shorter for larger densities.
Using some inert particles to decorate the water motion, we have noted that
not only do the animals move but also the water moves in the same direction
after the vortex is fully formed. Thus we call this a biological-hydrodynamic
vortex.  Below in Sect.~\ref{sec:outline-rwt}, we describe this phenomenon as
a physical phase transition. Within the framework of the RWT, we hypothesize
that the symmetry of the limit cycle pair is broken by a positive feedback
represented by a non-zero order parameter that evolves in time. If a momentary
fluctuation results in even a small majority of animals circling in one
direction with water being dragged along with them, then that direction
becomes dominant as the other animals feel the hydrodynamic flow.  Animals
initially moving counter to the dominant flow then change their direction thus
forming the vortex.

In Section~\ref{sec:how-symmetry-broken} below, we show that in order for the
vortex to form, some symmetry breaking mechanism must be present. We suggest
three possible processes: short range repulsion, avoidance and velocity
aligning hydrodynamic coupling. {\em Daphnia} do exhibit avoidance events as
discussed below in Sect.~\ref{sec:how-symmetry-broken}. It is important to
understand that avoidance is different from collisions between particles
mediated by a symmetric short-range repulsive potential.

\section{How is the symmetry broken?}\label{sec:how-symmetry-broken}

As discussed above, the two motions characteristic of small density
experiments with {\em Daphnia}, and described by the theory for
non-interacting particles, are the noisy fixed point followed by a bifurcation
to a pair of closed paths (limit cycles) that describe the equally probable
CCW and CW rotational motions in a central potential. Many particles with
attractive interactions can exhibit the third motion - swarming - this is
represented by collective motions toward a center together with the
aforementioned bi-directional rotations. The forth motion - the vortex - is
also a many-particle phenomenon but is characterized by the single rotational
motion of all the particles, and in the experiment, includes also fluid
rotation. As we have seen, all these motions are observable in our apparatus,
the former two with low {\em Daphnia} density and the latter two only in
experiments with large densities. We also have mentioned that theoretically
some symmetry breaking process is necessary in order to replace the two
counter rotating limit cycles with a single vortex rotating in a single
direction. The question is, in natural phenomena involving swimming animals,
what is the symmetry breaking force or process?

We put fourth two possibilities: short range repulsion and hydrodynamic
self-alignment of velocities. Perhaps the simplest theoretical
symmetry-breaking mechanism is offered by a symmetric short-range repulsive
potential between pairs of particles. The physical laws governing scattering
thus describe encounters among such particles. Avoidance is, however, not the
same as particle scattering. It involves animate behavior and arises, for
example, in dynamical descriptions of pedestrian avoidance maneuvers.
Hydrodynamic coupling involves the overlap of the flow field due to the motion
of one moving particle with that of another. As we show below, this coupling
tends to align velocities leading to a final rotational state in one
direction.

A third mechanism for symmetry breaking is collision avoidance. The first and
second are discussed below in some detail, while this is only outlined as it
has been described elsewhere. As we use the term here, avoidance has
behavioral implications and thus involves mechanisms that cannot be described
by simple, symmetric potentials \citep{MaOrSchw03} as, for example, the Morse
potential used below. We do not discuss avoidance in detail here. A similar
problem arises in dynamical descriptions of pedestrian avoidance maneuvers
\citep{HeMoSchw94}. These again lead to a symmetry breaking and consequently
to various collective motions including the vortex formations of interest
here.

\begin{figure}[htbp]
  \begin{center}
    \epsfig{file=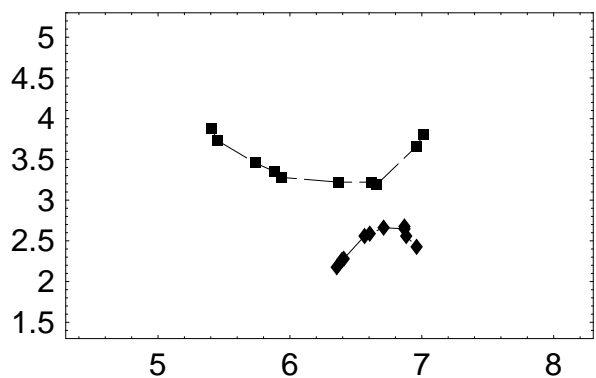,width=8cm}
    \caption{Track of a bottom view of an example avoidance event. The two 
      {\em Daphnia} are shown in their initial positions, being at the same
      height in the water column (as only a 6mm thick horizontal layer is
      illuminated by an IR laser and line generator). The tracking software is
      used to depict their trajectories through subsequent frames. In this
      example the {\em Daphnia} avoid each other, in a few other (rare)
      examples they actually collide.}
    \label{fig:13}
  \end{center}
\end{figure}

Both scattering due to a symmetric short-range repulsive potential and
hydrodynamic velocity alignment work equally well within the framework of the
ABP theory to break the limit cycle symmetry. Unfortunately, at present the
experiment does not provide enough information to distinguish between them.
Moreover, Fig.~\ref{fig:13}, showing an actual encounter of two {\em Daphnia},
suggests that they execute a collision avoidance maneuver.
Here two {\em Daphnia} are approaching each other in the horizontal plane,
avoid each other and depart on different trajectories. Though we have observed
numerous encounters between pairs of {\em Daphnia}, unfortunately, the
detailed statistics necessary for realistic modeling have not yet been
compiled.

\subsection{Short-range repulsion}

\begin{figure}[htbp]
  \begin{center}
    \epsfig{file=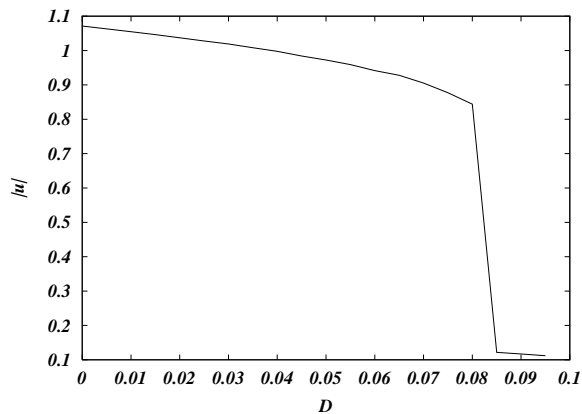,width=8cm}
    \caption{\label{fig:14} Noise induced transition from directed to
      motion with zero mean translational velocity but non zero
      mean angular momentum (short range repulsive potential -- Morse
      potential).}
  \end{center}
\end{figure}

Let us assume that the interaction between the particles shows long-range
attractive together with short-range repulsive behavior. As suggested by
similarities to interactions between molecules, we use the Morse-potential to
model this type of interaction. The resulting collective effect is the
formation of clusters, a well known effect from solid state physics
\citep{MoStu29,Mo29,Ki95}. In the limit of large inter-particle separation,
the Morse potential becomes approximately quadratic, that is, similar to the
external and mean field potentials used in
Sections~\ref{sec:outline-abp-theory} and \ref{sec:many-particle-abp}. As
those sections show, the self-propelling velocity together with a quadratic
potential lead to rotational motions (the two symmetric limit cycles). But
now, the short-range parts of the Morse potential parallelizes the neighbors and
breaks the symmetry of the limit cycles, leading to rotation of the entire
cluster in one (or the other) direction. As we know from the previous sections
rotations are stable within such potentials. Thus we can think of a cluster
with an initial condition such that all stationary velocities of the particles
are equal in magnitude and additionally all have the same direction.
Increasing the noise within the system leads to a phase transition as shown in
Fig.~\ref{fig:14}, where, for $D$ greater than a critical value, $D_{\rm
  crit}$, the cluster stops its directed motion. The net motion of the cluster
vanishes. Finally, this can lead to a rotation of the cluster.
In this new type of noise-induced phase transition the mean translational
velocity crashes to zero for $D \rightarrow D_{\rm crit}$ and is replaced by a
mean rotational velocity. The stability of the rotational motions was
discussed in detail in \citet{ErEbAn00} and the noise-induced transition in
\citet{ErEbMi01}.

\subsection{Hydrodynamics in the ABP theory}

\begin{figure}[htbp]
  \begin{center}
    \epsfig{file=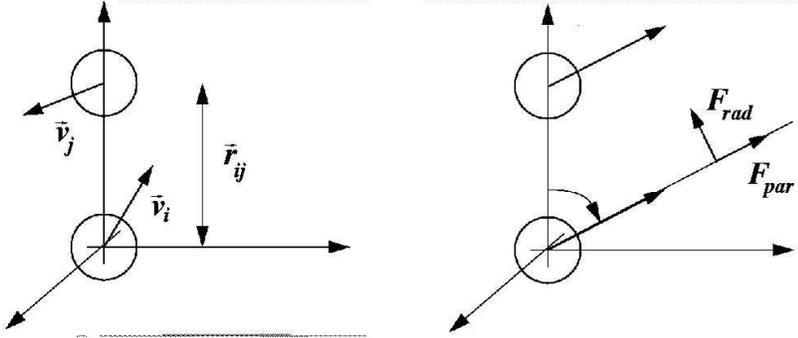,width=11cm}
    \caption{Geometry of the hydrodynamic coupling showing parallel and 
      perpendicular velocity components of the animals, before coupling
      through the flow field (left) and after coupling (right).
      $\vec{F}=\kappa_{\rm F}\vec{v}_{\rm F}$, where the vector velocities are
      defined in Eqs.~(\ref{eq:11})}
    \label{fig:15}
  \end{center}
\end{figure}

{\em Daphnia} live in a low Reynolds number fluid environment.  Hence, we can
imagine that laminar flows of water mediate the interaction between
individuals. The approach to use hydrodynamic interaction terms to model this
coupling has already been used before for some species
\citep{We73a,HuWi90,Ni94,Ni96a}. With this in mind, we consider the
dipole-like flow around one particle, $j$, that is moving at the
self-propelling velocity. We want to know the flow field of $j$ at a distance
$r_{ij}$ at the location of particle $i$, as shown in Fig.~\ref{fig:15}.
Considering similar contributions from all the other particles, a net
hydrodynamic velocity $v_{\rm F}$ is present at the location $i$.

In order to add a hydrodynamic interaction into Eq.~(\ref{eq:1}), this
stochastic equation of motion can be rewritten in the following way:
\begin{equation}
  \label{eq:9}
  m\partial_t\vec{v}_i=-\gamma(v_i)\,\vec{v}_i+\kappa_{\rm F}\,\vec{v}_{\rm F}
  -\nabla U(\vec{r}_i)+\sqrt{2D}\,\vec{\xi}_i(t)\,,
\end{equation}
where the term $\kappa_{\rm F}\,\vec{v}_{\rm F}$ is generated by all the other
moving particles, and the parameter, $\kappa_{\rm F}$ measures the strength of
the hydrodynamic interaction. This term tends to align the direction of the
velocity $\vec{v}$ with that of $\vec{v}_{\rm F}$. Thus particle $i$ is
finally dragged in the direction of motion of $j$. The aligning force, given
by the new term in Eq.~(\ref{eq:9}), originates with the viscosity of the
surrounding liquid, and can be modeled by a well-known term in the
Navier-Stokes equation \citep[see e.g.][]{LaLi_VI_engl}. Within the laminar
regime, and for low Reynolds number, the flow can be modeled by additive
Oseen-contributions \citep{Os15a,Os15b,Os15c}. This way of modeling
hydrodynamic interactions is similar to the one used in the theory of
electrolytes \citep{Fa71engl} and macromolecules
\citep{HuDo93,TiDa95,NeAdLyDa02}. Within this approximation the flow field
becomes
\begin{equation}
  \label{eq:10}
  \vec{v}_{\rm F}(\vec{r_i}) = \sum_{j\neq i} \frac{R}{r_{ij}}
  \left[\delta+\frac{\vec{r}_{ij}\otimes\vec{r}_{ij}}{r_{ij}^2}\right]\;
  \vec{v}_j
  \qquad {\rm valid~for~} r_{ij} \gg R\,,
\end{equation}
where $R$ is an effective hydrodynamic radius of the particles, and $\otimes$
is a tensor product. Strictly speaking the Oseen-law is valid only
asymptotically for $r_{ij} \gg R$. We do not claim that this is the only way
to introduce hydrodynamic interactions, nevertheless the Oseen-law has the
right symmetry properties. This is to be seen from the alternative way of
writing Eq.~(\ref{eq:10}):
\begin{equation}
  \label{eq:11}
  \vec{v}_{\rm F}(\vec{r_i}) =\sum_{j\neq i} \left[\frac{R}{r_{ij}}\; \vec{v}_j 
  + \frac{R\left(\vec{r}_{ij}\cdot\vec{v}_j\right)}{r_{ij}^3}\,\vec{r}_{ij}\right]
  \qquad {\rm valid~for~} r_{ij} \gg R\,.
\end{equation}
This shows that the Oseen-flow consists of parallel and radial components. The
first term on the right of Eq.~(\ref{eq:11}) is the component of the flow
field parallel to velocity of the particle $i$, and it is this component that
tends to align velocities of $i$ and $j$. The second term is the radial
component and tends to act along the direction of the line of sight between
$i$ and $j$. The alignment effect that results in symmetry breaking and thus
large-scale coherent motions is due to the parallel components.

If one imagines that we start with a population of active particles in a
liquid medium, confined by a quadratic potential, due to the combination of
active friction and external confinement on average half of the particles will
rotate CW and the other half CCW. In contrast to particles that move in a
vacuum, coupling is mediated through the liquid. This leads to an interaction
as described, for example, by Eq.~(\ref{eq:11}). Because the hydrodynamic
interaction aligns the velocity vectors of the individuals the dual rotational
motion breaks down and the final motion is of the entire population rotating
in one direction. This direction is determined by random initial conditions.
Additional motions, including clustering, global rotation of clusters and
final rotational direction selection, result from this hydrodynamic
interaction. These have been described in detail in \citep{ErEb02}.

\section{Random Walk Theory (RWT)}\label{sec:outline-rwt}

\subsection{Introduction}
\label{sec:intro_RWT}

Discrete Random Walk Theories (RWTs) have been used extensively in biology,
mainly to describe the irregular motions of bacteria \citep{Be93}. RWTs are
iterative, that is a certain set of rules is applied to a particle located at
a certain point in space in order to compute its future location and this
process is repeated (iterated) to compute sets of jumps to successive
locations. Since {\em Daphnia} naturally move in sequences of hops separated
by pauses with a new direction of motion chosen at the end of each hop (see
Fig.~\ref{fig:6}), RWTs can generate descriptive models of their motions. In
this section we describe such a model and compare its predictions with
experimental data. The experimental results reported in
Section~\ref{sec:exper-with-small} for {\em Daphnia} at low densities indicate
that the dominant interaction is with the central attractant and that
inter-animal interactions are much less important for low animal densities.
Consequently theoretical models of self-propelled, interacting animals and
their variants \citep{ViCzBeCoSh95,LeRaCo00,CoKrJaRuFr02} are not applicable
to experiments at low densities with our animals. Though the ABP theory
outlined in Section~\ref{sec:outline-abp-theory} adequately explains the noisy
fixed point and limit cycle motions, we outline here a theory based on random
walks. This theory has the advantage of being simple and easily simulated
numerically, since it is only necessary to iterate and update an algorithm
rather than to integrate a set of stochastic differential equations.

Any theory of {\em Daphnia} motion at low densities within an attracting field
must predict the two observed motions: noisy movements near and through a
fixed point and the rotational motions about the attractant.  An obvious
question arises: what are the minimum ingredients necessary to describe these
motions?  To answer this question, and to simulate the observed behavior of
single {\em Daphnia}, we developed a self-propelled particle model based on
random walks with the aim of being as simple as possible. The final model for
low animal densities, closely related to the ABP theory, consists of only two
ingredients: a short-range temporal correlation to simulate the general
movement of the {\em Daphnia} in darkness and an attraction to the light
marker \citep{OrBaMo02a}.

\subsection{The Turning Angle Distribution}
\label{sec:turning}

\begin{figure}[htbp]
  \begin{center}
    \begin{center}
      \unitlength 0.3mm
      \def\epsfsize#1#2{0.36#1}
      \begin{picture}(250,170)
        \put(0,15){\epsfbox{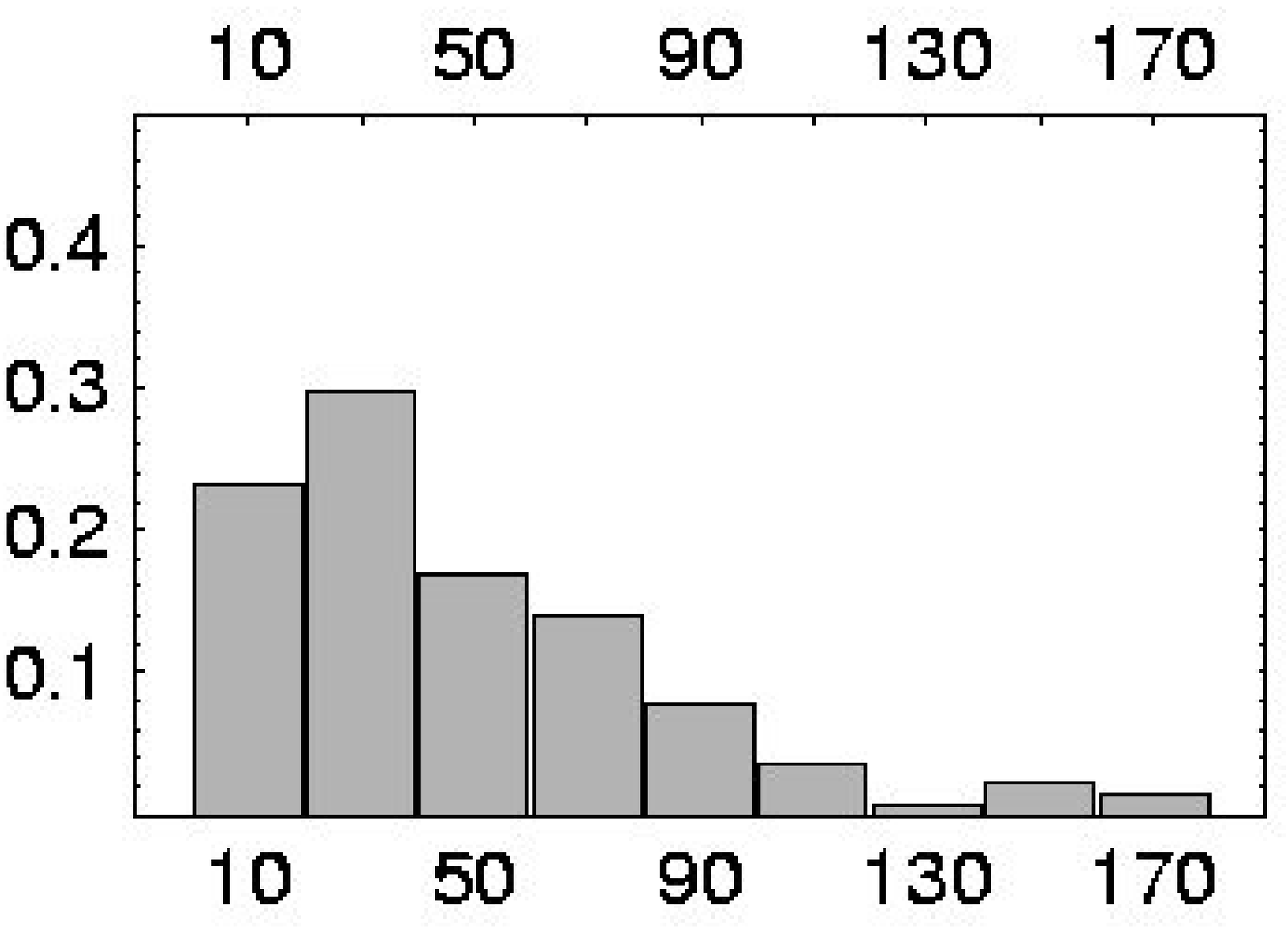}}
        \put(23,148){\epsfig{file=fig/09/whitespace.eps,height=5mm,width=70mm}}
        \put(-2,78){\makebox(10,10)[r]{\rotate[l]{\fontsize{15}{15}
              \selectfont{$\displaystyle P(| \alpha |)$}}}} 
        \put(128,0){\makebox(10,10)[r]{{{\fontsize{15}{15}
                \selectfont{$\displaystyle  | \alpha | $}}}} }
      \end{picture} 
    \end{center}
    \caption{The experimentally measured distribution of absolute turning
      angles, $|\alpha|$. The histogram was assembled from analysis of 1599
      hops taken from eight animals moving freely in a dark aquarium with
      uniformly distributed food. The {\em Daphnia} were conditioned in the
      dark for a minimum of 15 min. before recording data. The turning angles
      were projected onto the two-dimensional horizontal plane. Plot adapted
      from \citet{OrBaMo02a}.}
    \label{fig:16}
  \end{center}
\end{figure}

Discrete random walk models having memory, or correlation, are powerful tools
to simulate various biological, chemical and physical processes that are more
complex than pure Brownian motion \citep{Hu95}. In the case of our {\em
  Daphnia}, a short-range correlation is indicated by the fact that the
animal's choice of turning angle, $\alpha$, at the end of each hop is not
completely random but instead is correlated to some degree with the previous
direction.  We have observed a large number of hops of {\em Daphnia} swimming
in the dark.  From these data we have assembled the distribution $P(\alpha)$
of turning angles (DTA), which turns out to be a bimodal symmetric
distribution with a pair of maxima at approximately $\pm$35{\tc\symbol{176}}
\citep{OrBaMo02a}.  Shown in Fig.~\ref{fig:16} is the distribution
$P(|\alpha|)$ of the absolute turning angles, which resembles the results
found for the DTA of the oceanic zooplankton copepod \citep{SchmSe01}. The
presence of these maxima signals the existence of a short-range temporal
correlation. If the directions of motion were completely uncorrelated, as for
purely Brownian walkers, the DTA would be uniform. Thus {\em Daphnia} prefer
moving in the forward direction and the most probable turning angles
$\alpha_{\rm max} \cong \pm 35${\tc\symbol{176}} may have evolved to optimize
the amount of territory covered while foraging in patchy food distributions
\citep{GeStr77}.

In constructing our RWT we introduce the random process (noise) by randomly
choosing a new turning angle, $\alpha$, from the experimentally measured DTA
shown in Fig.~\ref{fig:16} for every iteration, or hop.
This represents a temporal correlation of only one iteration (time step).
Though {\em Daphnia} may have a memory extending over a few hops, our
observations indicate that this does not dominate the motion as the first hop
memory does. Moreover, including only a one step memory keeps the model
simple, which is our major aim. The particle in our RWT hops a fixed distance,
$\Delta x$, in a time step, $\Delta t$, upon each iteration. Thus the
magnitude of its velocity is a constant, and this is the self-propelling
velocity. These model conditions are realistic, since the hop length and mean
velocity of {\em Daphnia} are observed to be approximately constant
\citep{Do96}.

\subsection{Motions Stimulated by Light}

\begin{figure}[htbp]
  \begin{center}
    \epsfig{file=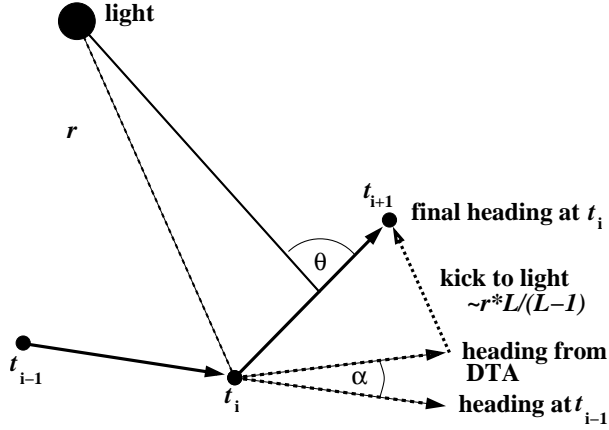,width=8cm}
    \caption{The geometry of a pair of iterations, $t_i$ and $t_{i+1}$, of 
      the RWT showing distance to the light source $r$, heading angle
      $\theta$, and turning angle $\alpha$, as taken from the DTA shown in
      Fig.~\ref{fig:16}.}
    \label{fig:17}
  \end{center}
\end{figure}

In order to simulate the behavior of {\em Daphnia} in a light field an
attraction to the marker must be built into the theory. The ABP theory
detailed above models the attraction with a linear restoring force in the
direction of the center of motion. Similarly, we construct a "kick"
proportional to the distance, $r$, of the particle to the light source and
proportional to a normalized strength parameter $L/(1-L)$, where $[0 \leq L <
1]$. For every time step, the walker randomly chooses a directional change,
$\alpha$, from the DTA, then the kick, $[L/(1-L)]r$ towards the light is
applied resulting in a new heading angle $\theta$. The final heading vector,
labeled $t_{i+1} - t_i$, is rescaled to unit length to maintain constant
velocity of the walker. The geometry of a pair of iterations for the time
steps $\Delta t = t_i - t_{i-1}$ and $\Delta t = t_{i+1} - t_i$ is shown in
Fig.~\ref{fig:17}.

With the RWT outlined above, the moves of the walker can be simulated for
varying attraction strength $L$. For $L \cong 0.4$ the walker revolves around
the light source in both directions and frequently changes its rotational
direction similar to the recorded {\em Daphnia} motions shown in
Fig.~\ref{fig:8}.  Thus the RWT predicts limit cycle motions similar to those
of the ABP theory.  For smaller values of $L$, the circular motion is less
pronounced as the influence of the randomness increases. For larger $L$, the
circling breaks down and the walker mainly steps back and forth over the light
source. This simulates the noisy fixed point motions similar to those
predicted by the ABP theory. But in the RWT, the strength of attraction to
light governs the type of movement.  That is, in the RWT, the shape of the
potential varies while the self-propelling velocity is constant. Whereas, in
the ABP theory, the shape of the potential is fixed while the self-propelling
velocity varies (this velocity is determined by the parameters of the energy
depot and the friction coefficient).

\begin{figure}[htbp]
  \begin{center}
    \hspace*{-2cm}
    \subfigure[]{
      \unitlength 0.15mm
      \def\epsfsize#1#2{0.34#1}
      \begin{picture}(300,230)
        \put(0,12){\epsfbox{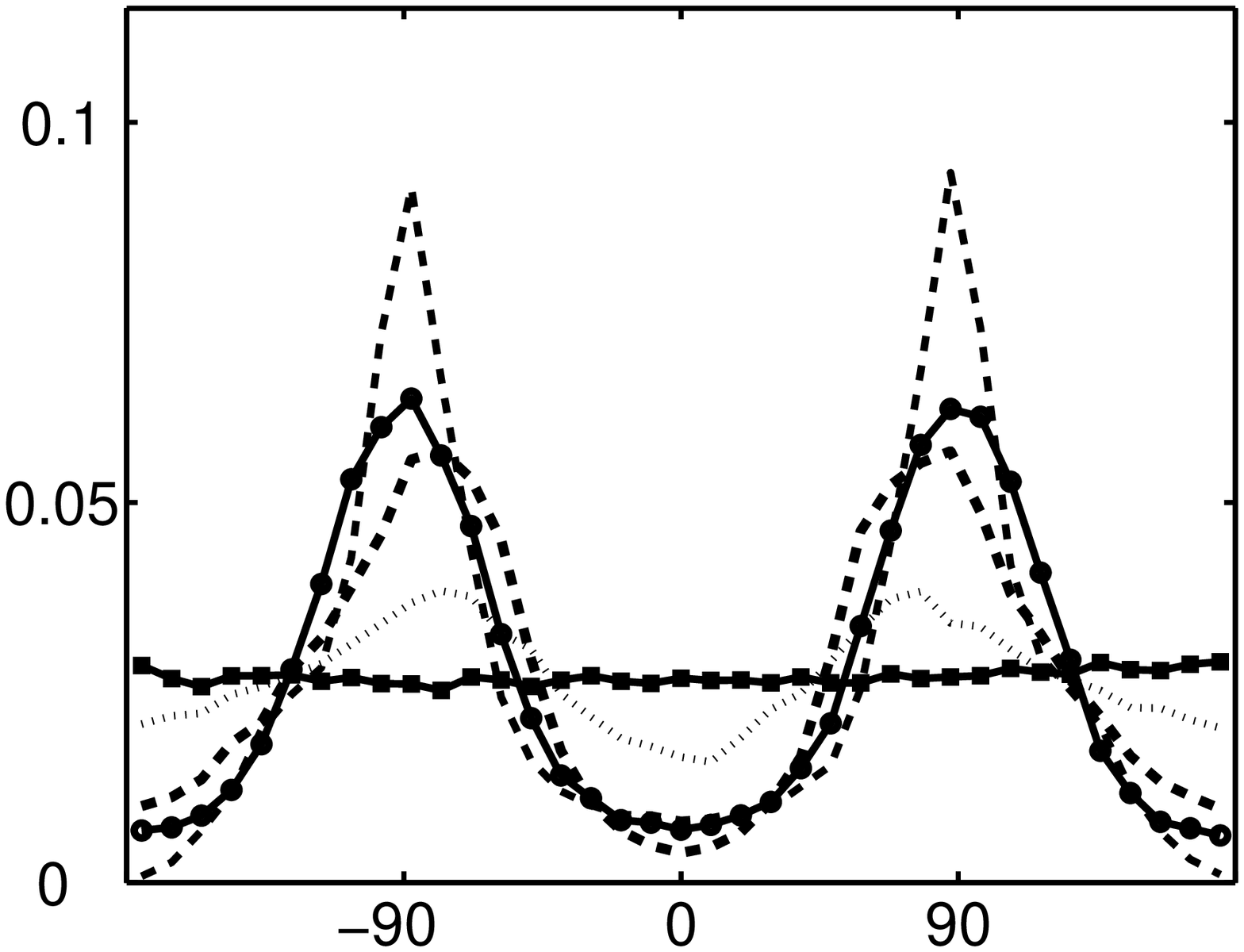}}
        \put(-18,172){\makebox(10,10)[r]{\rotate[l]{\fontsize{12}{12}
              \selectfont{$\displaystyle P(\theta)$}}}} 
        \put(150,0){\makebox(10,10)[r]{{{\fontsize{12}{12}
                \selectfont{$\displaystyle \theta$}}}} }
      \end{picture} 
      \label{fig:18a}}
    \hspace*{2cm}
    \subfigure[]{
      \unitlength 0.15mm
      \begin{picture}(300,230)
        \def\epsfsize#1#2{0.34#1}
        \put(20,12){\epsfbox{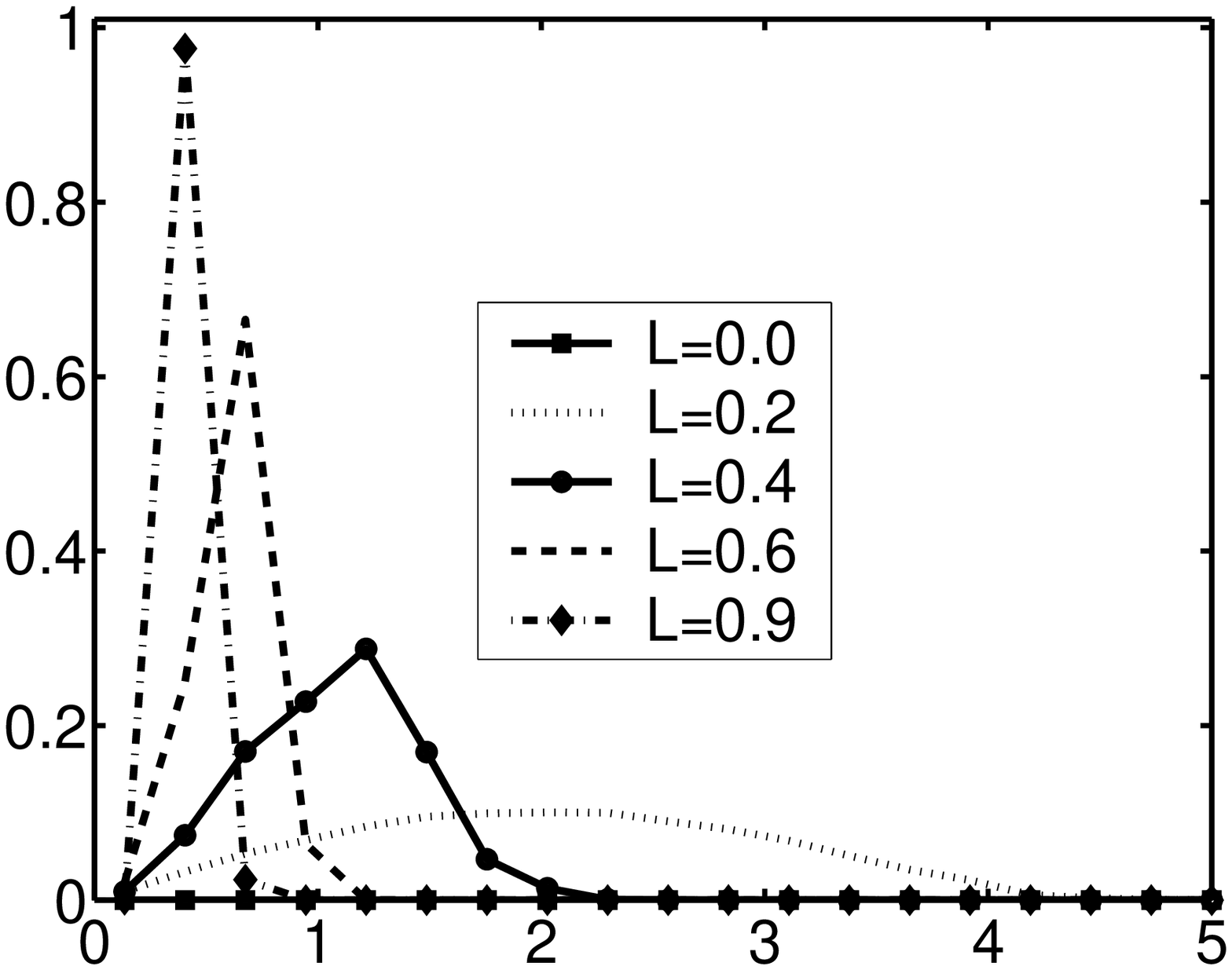}}
        \put(-18,172){\makebox(10,10)[r]{\rotate[l]{\fontsize{12}{12}
              \selectfont{$\displaystyle P(r)$}}}} 
        \put(150,0){\makebox(10,10)[r]{{{\fontsize{12}{12}
                \selectfont{$\displaystyle r$}}}} }
      \end{picture} 
      \label{fig:18b}}\\[1.2cm]
    \hspace*{-2cm}
    \subfigure[]{
      \unitlength 0.15mm
      \begin{picture}(300,230)
        \def\epsfsize#1#2{0.34#1}
        \put(0,12){\epsfbox{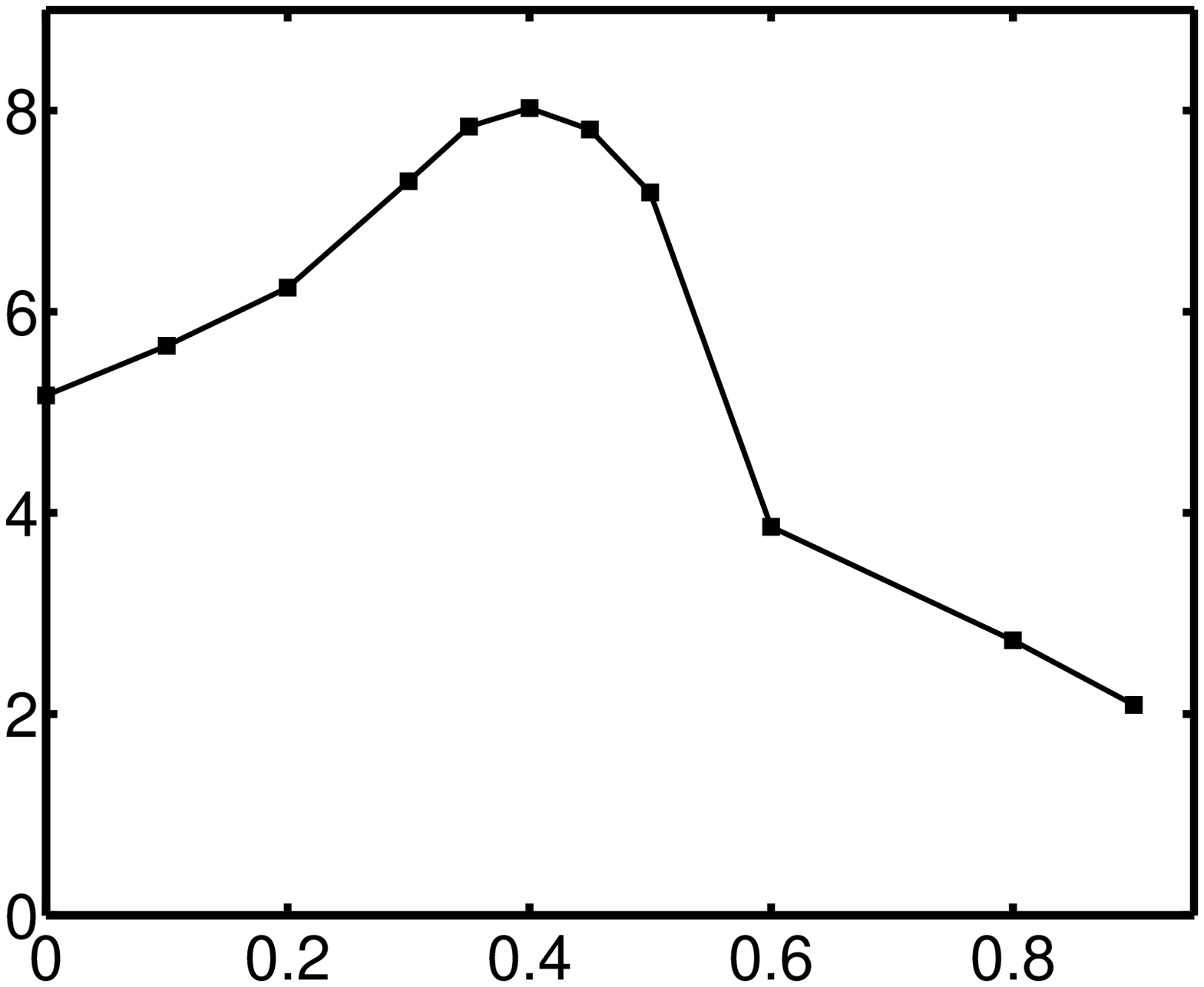}}
        \put(-18,172){\makebox(10,10)[r]{\rotate[l]{\fontsize{12}{12}
              \selectfont{$\displaystyle \left<M_{C}\right>$}}}} 
        \put(195,0){\makebox(10,10)[r]{{{\fontsize{12}{12}
                \selectfont{$\displaystyle  L $}}}} }
      \end{picture} 
      \label{fig:18c}}
    \hspace*{2cm}
    \subfigure[]{
      \unitlength 0.15mm
      \def\epsfsize#1#2{0.34#1}
      \begin{picture}(300,230)
        \put(0,12){\epsfbox{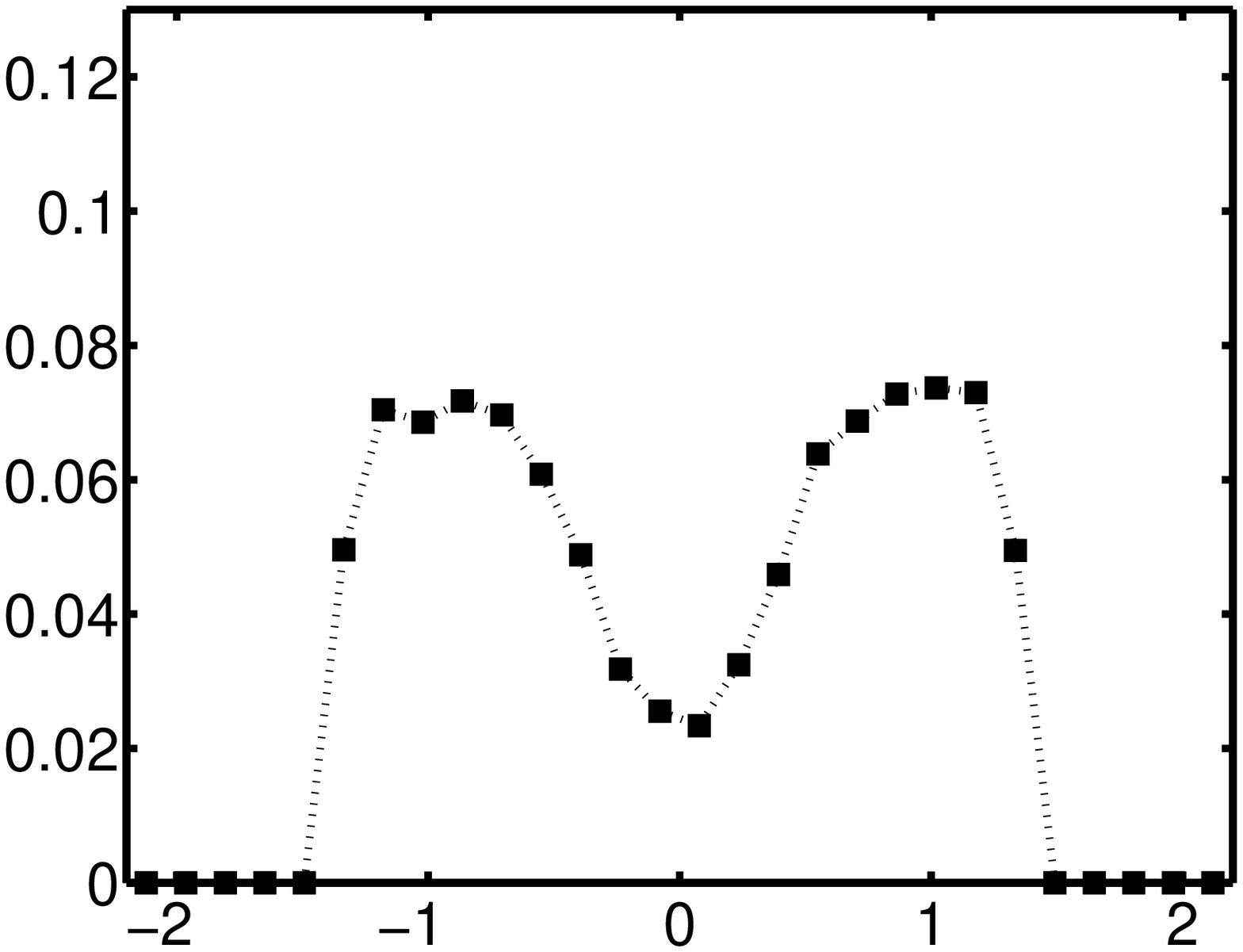}}
        \put(-10,172){\makebox(10,10)[r]{\rotate[l]{\fontsize{12}{12}
              \selectfont{$\displaystyle P(L_{\mathrm{ang}})$}}}} 
        \put(240,0){\makebox(10,10)[r]{{{\fontsize{12}{12}
                \selectfont{$\displaystyle  L_{\mathrm{ang}}$}}}} }
      \end{picture} 
      \label{fig:18d}}\\[1.2cm]
    \subfigure[]{
      \unitlength 0.15mm
      \def\epsfsize#1#2{0.34#1}
      \begin{picture}(300,230)
        \put(0,12){\epsfbox{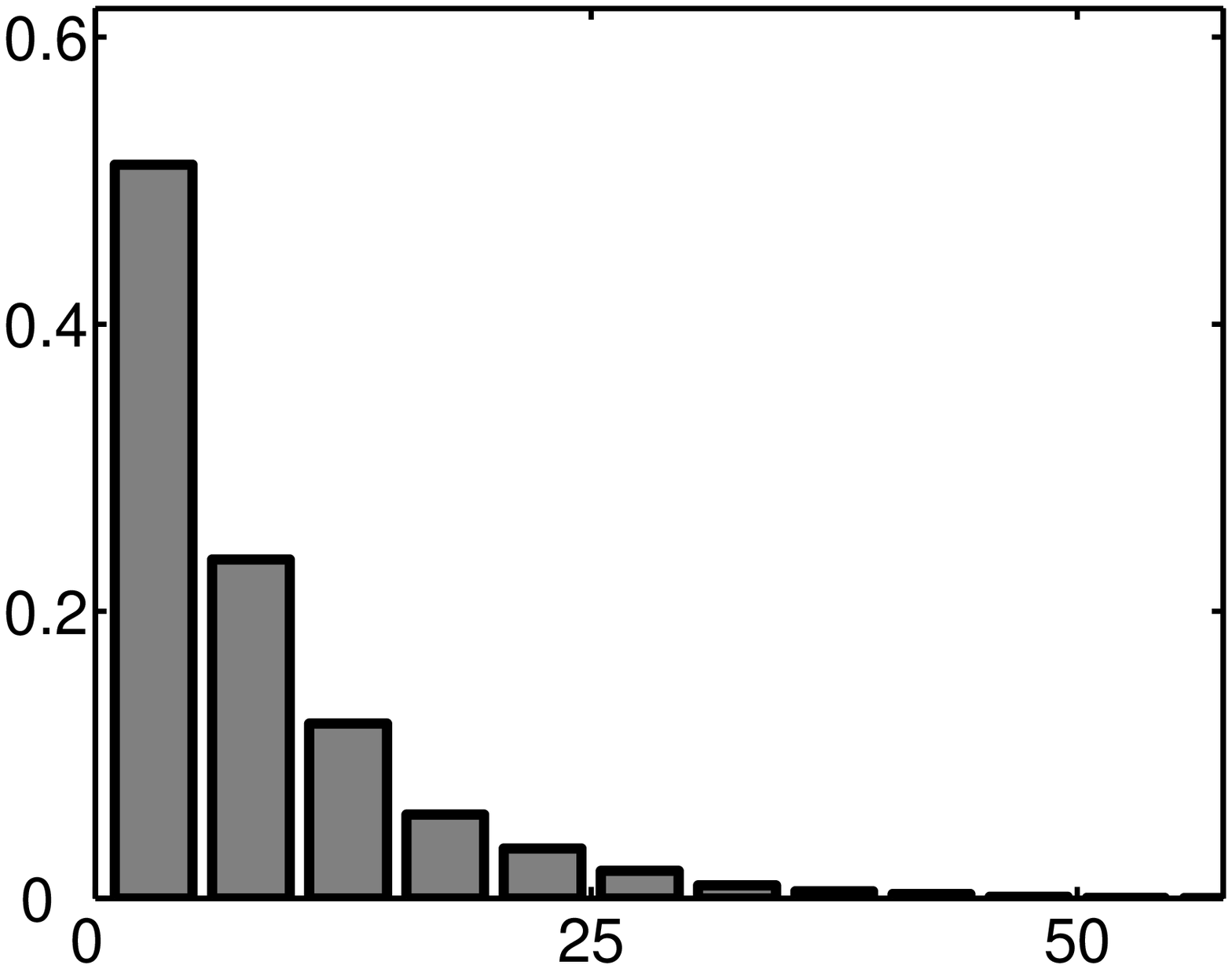}}
        \put(-18,172){\makebox(10,10)[r]{\rotate[l]{\fontsize{12}{12}
              \selectfont{$\displaystyle P(M_{C})$}}}} 
        \put(240,0){\makebox(10,10)[r]{{{\fontsize{12}{12}
                \selectfont{$\displaystyle M_{C}$}}}} }
      \end{picture} 
      \label{fig:18e}}
    \caption{Statistical measures predicted by the RWT for $L= 0.0$ (black 
      squares), 0.2 (dotted line), 0.4 (black circles), 0.6 (dashed line), and
      0.9 (black diamonds), see also inset of (b). (a) The distribution of
      heading angles, $P(\theta)$. The twin maxima at $\pm$90{\tc\symbol{176}}
      demonstrate equally probable rotational motions in both directions (CW
      and CCW). (b) The distribution of radii, $P(r)$. (c) The average number
      of iterations $\langle M_{\rm C}\rangle$ before a change in the circling
      direction takes place for different $L$. Note the maximum at $L=0.4$
      indicating the case of most pronounced circling. (d) The distribution of
      angular momentum, $P(L_{\rm ang})$, and (e) the distribution of the
      number of hops before a change of rotational direction, $P(M_{\rm C})$,
      both at $L=0.4$. Plots are adapted from \citet{OrBaMo02a,OrBaMo02b}.}
    \label{fig:18}
  \end{center}
\end{figure}

We turn now to measures of average quantities as predicted by the RWT. As in
the experiment discussed in Section~\ref{sec:exper-with-small}, rotational
motion is demonstrated by the distribution of heading angles, $P(\theta)$. The
model predicts a similar distribution with equally probable rotational motions
in both directions as shown in Fig.~\ref{fig:18a} for several values of the
light strength, $L$. The distribution of radii, $P(r)$, is shown in
Fig.~\ref{fig:18b} again for the same set of values of $L$.
Fig.~\ref{fig:18c} displays the average number of iterations $\langle M_{\rm
  C}\rangle$ before a change in the circling direction for different $L$. A
maximum at $L=0.4$ implies that the most pronounced circling takes place at
this intermediate value of the attraction strength. Figure~\ref{fig:18d} shows
the angular momentum distribution, $P(L_{\rm ang})$ for this optimal value of
$L$. Finally, the distribution $P(M_{\rm C})$ of the number of hops
(iterations) before a reversal of direction is shown in Fig.~\ref{fig:18e},
again for $L=0.4$. For comparison of the simulation results to the
experimental data see the figures in Section~\ref{sec:exper-with-small}. A
more detailed discussion of these results can be found in \citet{OrBaMo02a}.
%\enlargethispage{1cm}

\subsection{Diffusion in RWT I}
\label{sec:diff_1}

\begin{figure}[htbp]
  \begin{center}
    \unitlength 0.3mm
    \def\epsfsize#1#2{0.715#1}
    \begin{picture}(250,170)
      \put(8,15){\epsfbox{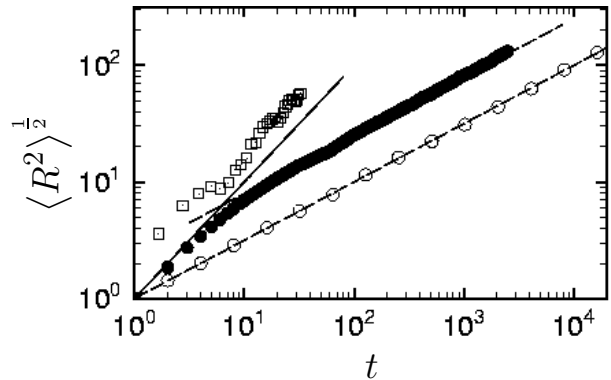}}
      \put(0,85){\makebox(10,10)[r]{\rotate[l]{\fontsize{15}{15}
            \selectfont{
              $\displaystyle \left\langle R^2\right\rangle^{\frac 12} $}}}} 
      \put(136,0){\makebox(10,10)[r]{{{\fontsize{15}{15}
              \selectfont{$\displaystyle  t $}}}} }
    \end{picture} 
    \caption{Diffusion of {\em Daphnia} and of random walkers with and without
      short time correlations. The root mean square (rms) end-to-end distance
      $\left\langle R^2\right\rangle^{\frac 12}$ is plotted versus time, $t$.
      For the simulations, 500 realizations were used with a one step
      correlation (full circles) and zero correlation (open circles). The
      experimental data (open squares) was obtained from the trajectories of
      eight {\em Daphnia}, where the distances are in mm and the time is in
      sec. The slope of the dashed line is $1$ and of the solid lines $1/2$,
      indicating ballistic (correlated) and uncorrelated motions,
      respectively.}
    \label{fig:19}
  \end{center}
\end{figure}

We can study the natural diffusion of {\em Daphnia} by once again considering
their motions in the dark. Within the framework of the RWT, a convenient
measure of diffusion with time, is the end-to-end distance $R$. This is the
straight-line (Euclidean) distance from the starting point of a walkers
trajectory to the end point at the time $t_i$. Of course, an ensemble average
of such distances must be obtained from a large number of trajectories, each
generated with a different seed for the random number generator. Of interest
is the exponent, $\nu$, defined by $\left\langle R^2\right\rangle^{\frac
  12}\propto t^{\nu}$, where $t$ is the time (or the number of iterations)
after the walker is started, and the angular brackets indicate the
aforementioned ensemble average \citep[see for example][]{Hu95}. For purely
Brownian particles, that is those with zero correlation, $\nu = 1/2$.  In
contrast, for our particles with a temporal short-range positive correlation,
we find $\nu = 1$ for short time scales, followed by a crossover to $\nu =
1/2$ for asymptotically long times. This is shown by the simulation results
(full circles) in Fig.~\ref{fig:19}.
In general, for short-range temporal correlations, the crossover time from
non-Brownian motion to Brownian motion depends on the magnitude of the
correlation time (memory). Here the correlation time is just one time step,
since we simply update the turning angle according to the DTA at each
iteration. For comparison, we show a simulation for purely Brownian walkers
(open circles in Fig.~\ref{fig:19}) for which $\nu = 1/2$ always. When
calculating the rms end-to-end distance of {\em Daphnia} motions in the dark,
we observe an exponent $\nu\cong 1$ for $t\leq 30{\rm sec}$ (see squares in
Fig.~\ref{fig:19}). Thus their motions are governed by positive correlations at
least for short time scales \citep[for more details see][]{OrBaMo02b}. It is
widely assumed that for long time scales, that require much larger water
reservoirs, the tracks of zooplankton in the field resemble Brownian motion,
in agreement with our simulation at long times.  Unfortunately, we were unable
to observe the crossover experimentally, because we cannot record long enough
trajectories. The trajectory length was limited by the size of our aquarium
and by the available field of view in respect to the resolution of the video
cameras.

\subsection{Diffusion in RWT II}
\label{sec:diff_2}

As has been observed in the experiments {\em Daphnia} tend to turn with a
preferred angle at every hop they make (see Fig.~\ref{fig:16}). The question
which occurs, having this in mind, is: Why do {\em Daphnia} hop with an DTA as
shown above? What could be an reason for it? Can it be an evolutionary one? In
order to answer this question it is reasonable to calculate the diffusion
coefficient, having the aforementioned DTA in mind, and compare this with
normal Brownian motion as it has been done in \citet{KoErSchi03}. In order to
calculate the diffusion coefficient we derive the mean square displacement of
a swarm after $n=t/\Delta t$ time steps:
\begin{eqnarray}
  \left<R^{2}_n \right>
  = n\Delta x^2 + 2\Delta x^2 \sum_{i=1}^{n-1}\sum_{j>1}^n 
  \left<\cos{\alpha}\right>\,,
  \label{taylor-kubo}
\end{eqnarray}
which depends on the angular correlation function:
\begin{equation}
  \label{eq:ang_corr}
  C_{\alpha} =\left<\cos \alpha\right> 
  =\int_{-\pi}^{\pi}P(\alpha)\cos{\alpha}\,d\alpha\,.
\end{equation}
After some transformations \citep{OkLe02,Wu2000} of Eq.~(\ref{taylor-kubo})
the mean squared displacement of a two-dimensional correlated random walk can
be calculated what leads to the diffusion coefficient \citep{KoErSchi03}:
\begin{equation}
  4\,D  =\frac{1+C_{\alpha}}{1-C_{\alpha}}\,\frac{\Delta x^2}{\Delta t} \,.
  \label{D_gamma}
\end{equation}

\begin{figure}
  \begin{center}
    \subfigure[Angular correlation]{\label{gammagrafik}
      \epsfig{file=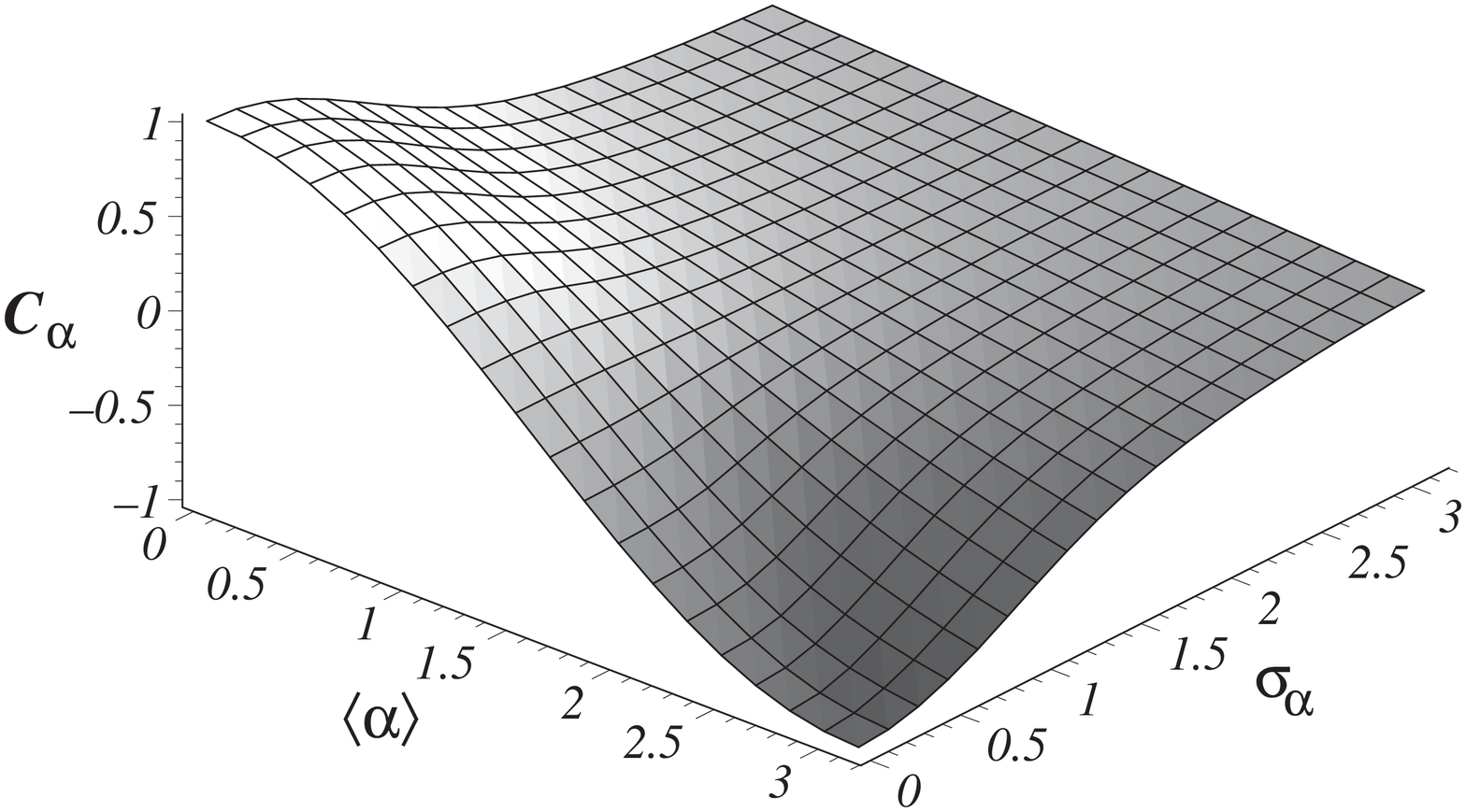,width=6.05cm}
      }
    \subfigure[Diffusion coefficient]{\label{d_koeff}
      \epsfig{file=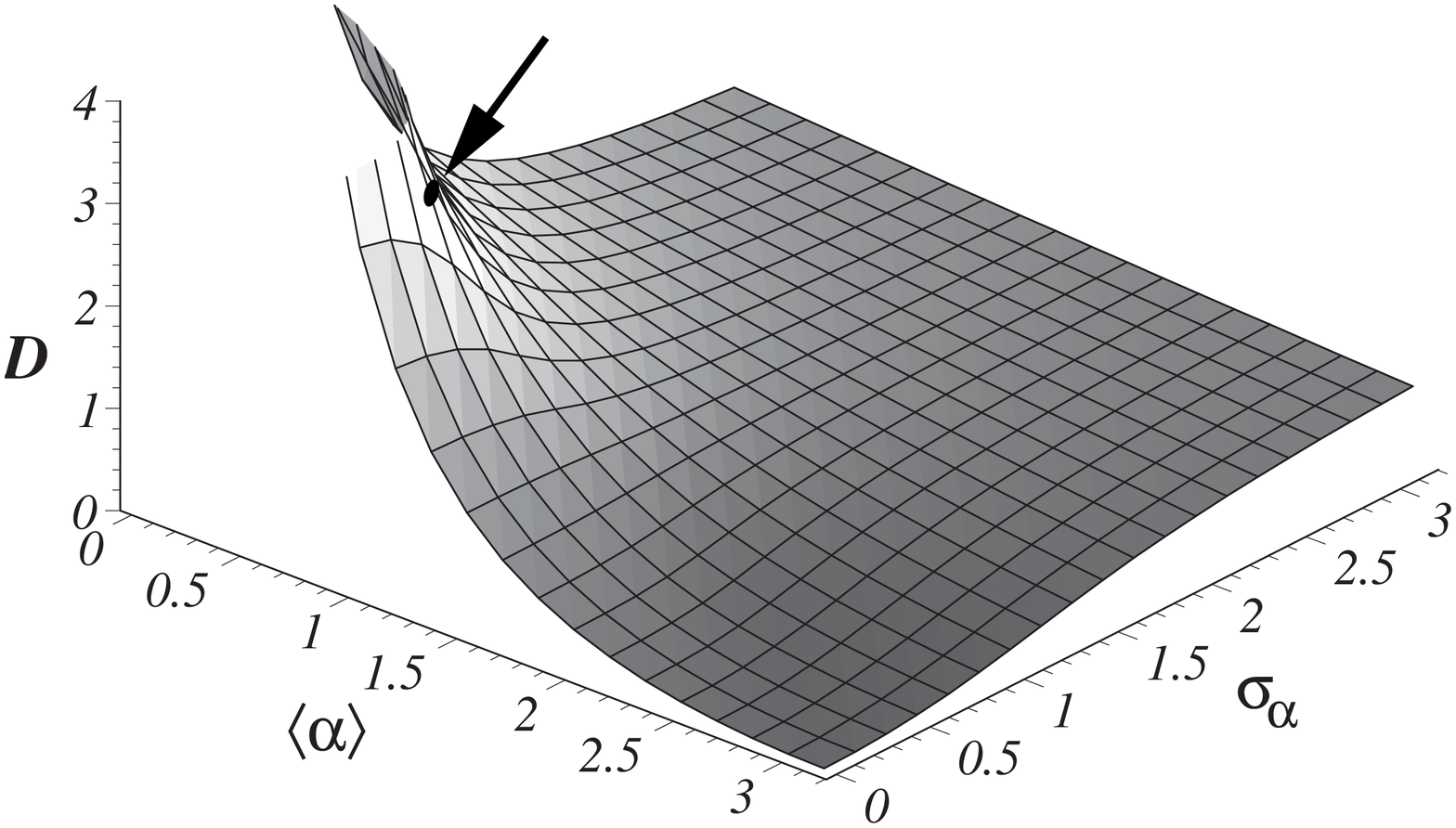,width=6.05cm}
    }
    \caption{\label{fig:21} (a) Angular correlation versus mean
      $\left<\alpha\right>$ and variance $\sigma_{\alpha}$ of a Gaussian like
      DTA. (b) Diffusion coefficient. The parameters out of the {\em Daphnia}
      experiments are indicated with an arrow at the following values:
      $\langle \alpha \rangle= 0.84$, $\sigma_{\alpha}=0.63$ and $D=3.1$.}
  \end{center}
\end{figure}

As Eqs.~(\ref{taylor-kubo}) and (\ref{D_gamma}) show the angular correlation
function is crucial for the behavior of the time evolution of the mean square
displacement of the particles and therefore for the diffusion behavior. If
$0<C_{\alpha}<1$ the diffusion would be enhanced. This could be compared with
a mean turning angle $0<\alpha<\pi/2$, as observed with the {\em Daphnia}. In
\cite{KoErSchi03} $C_{\alpha}$ has been calculated analytically assuming the
DTA can be approximated by a Gaussian distribution (shown in
Fig.~\ref{fig:21}).
For $\langle\alpha\rangle=\pi/2$ we have a $C_{\alpha}=0$ and therefore
$D=1$, what is equivalent to free diffusion, and does not depend on the
standard deviation of the turning angle anymore.

Taking the experimental parameters for the {\em Daphnia} looking at
Fig.~\ref{d_koeff} it seems to be obvious that the DTA for zooplankton has
evolutionary advantages. The diffusion is enhanced and therefore an animal is
able to cover a larger area while foraging for food.

\subsection{Breaking the Symmetry in the RWT}

We now turn to a simulation of the transition to vortex viewed as a phase
transition within the context of our RWT model. The model is modified based on
ideas about hydrodynamic feedback. We assume that the {\em Daphnia} are
sensitive to small water movements, and that they have a tendency to align the
direction of their motions with that of the majority of their neighbors. We
thus define a global order parameter, $O = (N_{\rm CW} - N_{\rm CCW})/N_{\rm
  tot}$ , where $N_{\rm CW}$ and $N_{\rm CCW}$ are the instantaneous number of
particles moving CW and CCW, respectively, and $N_{\rm tot}$ is the total
number of particles. Particles are considered to be executing rotational
motion if the absolute value of their heading angles lies within the range,
$80\leq |\theta|\leq 100$. In the simulation, $O$ evolves with time. It is
zero at instants when there is no net directional motion of the particles,
smaller or larger than zero when there is net CCW or CW motion respectively.
For the fully formed vortex (all particles moving in the same direction), $O
\rightarrow \pm 1$.

\begin{figure}[htbp]
  \begin{center}
    \subfigure[]{
      \unitlength 0.2mm
      \def\epsfsize#1#2{0.34#1}
      \begin{picture}(300,250)
        \put(0,12){\epsfbox{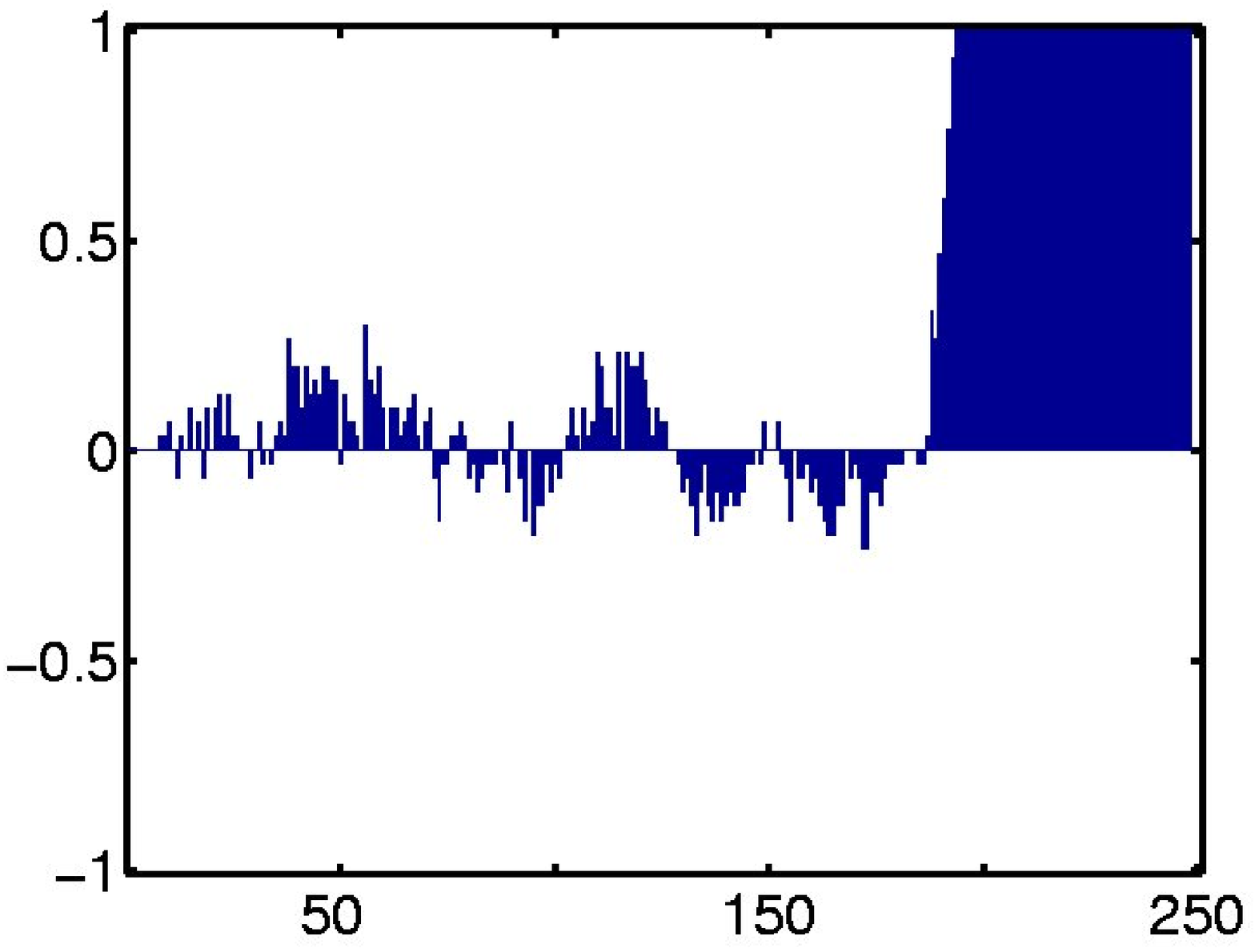}}
        \put(-10,150){\makebox(10,10)[r]{{\fontsize{15}{15}
              \selectfont{$\displaystyle O$}}}} 
        \put(150,0){\makebox(10,10)[r]{{{\fontsize{15}{15}
                \selectfont{$\displaystyle  t$}}}} }
      \end{picture} 
      \label{fig:20a}}
    \subfigure[]{
      \unitlength 0.2mm
      \def\epsfsize#1#2{0.34#1}
      \begin{picture}(300,250)
        \put(0,12){\epsfbox{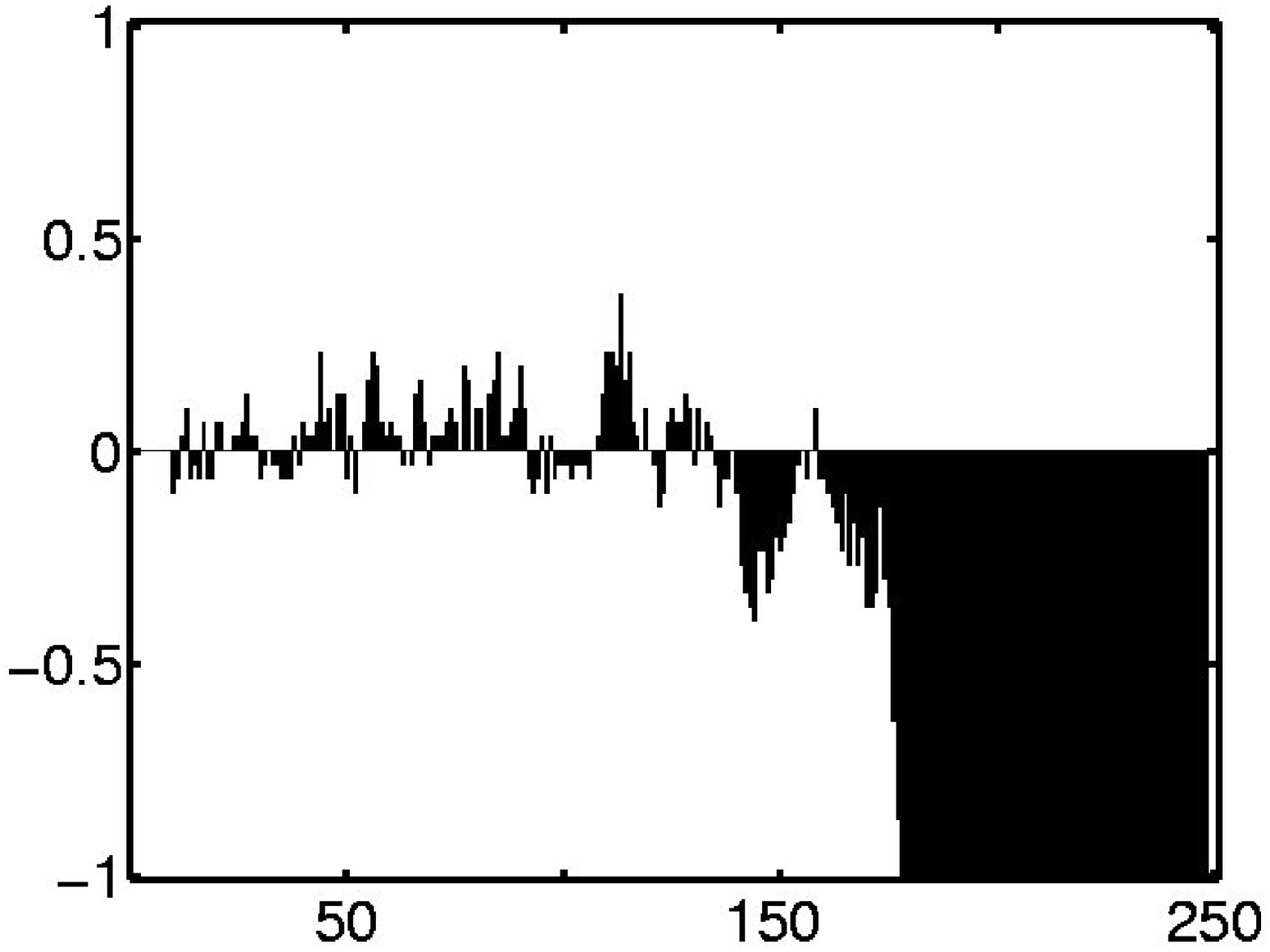}}
%         \put(-5,150){\makebox(10,10)[r]{{\fontsize{15}{15}
%               \selectfont{$\displaystyle O$}}}} 
        \put(160,0){\makebox(10,10)[r]{{{\fontsize{15}{15}
                \selectfont{$\displaystyle t$}}}} }
      \end{picture} 
      \label{fig:20b}}
    \caption{\label{fig:20} Evolution of the global order parameter, $O$,
      showing the formation of the vortex (a) in CW direction ($O \rightarrow
      +1$) at $t \simeq 190$ time steps for parameters, $N_{\rm tot} = 30$, $a
      = 0.7$, and $q = 0.16$, (b) in CCW direction ($O \rightarrow -1$) at
      $t \simeq 180$ time steps for the same parameters.}
  \end{center}
\end{figure}

We now focus on motions in a small neighborhood of radius $a$, supposing that
a single {\em Daphnia} can sense hydrodynamic disturbances only from its
neighbors.  Motions within the neighborhood of a particle are governed by a
local order parameter, $V = (A_{\rm CW} - A_{\rm CCW})/N_a$ , where $A_{\rm
  CCW}$ and $A_{\rm CW}$ are the numbers of particles within the neighborhood
that satisfy the same definition of rotational motion, and $N_a$ is the number
of all particles within the neighborhood. In addition to the kick toward the
light that applies to all particles in the simulation, we now include a local
kick, $K = qV$ in the appropriate direction (CW for $V > 0$, CCW for $V < 0$)
to the corresponding particle. Depending on the strength, $q$, of the kick and
the size of the neighborhood, a vortex forms after a certain time. The
formation of the vortex can be tracked by observing the temporal evolution of
the global order parameter $O$. An example is shown in Fig.~\ref{fig:20}.

\section{Discussion and Conclusions}

{\em Daphnia} provide for the first time, the opportunity to construct
well-controlled laboratory experiments with real biological agents. Such
experiments provide crucial tests of existing different theoretical models of
swarming, fixed point and rotational motions and vortex formation. And
therefore these experiments open up the possibility to learn more about the
aforementioned four different animal and agent motions through the interplay
of theoretical modeling and experimental investigations. For {\em Daphnia},
the interaction that leads to swarming and vortex formation is not a direct
one, as is the case for birds and fish that almost certainly use visual
velocity alignment.  Moreover, {\em Daphnia} swarms do not perform obvious
predator avoidance maneuvers such as swarm splitting and recombining
\citep{HaWaMa86}. Nevertheless, the observation of vortex-swarming by {\em
Daphnia} in the laboratory provides the possibility to learn more about the
physical, chemical and biological aspects of this phenomenon that may be
important for understanding similar motions exhibited by other living
creatures. Further experiments with {\em Daphnia} will be necessary to
understand in detail the factors that influence vortex-swarming, for example,
{\em Daphnia} species, predator kairomone concentration, density of both food
and individuals, and light intensity and wavelength. Light patterns and their
perception by {\em Daphnia} as well as the physical aspects of the
bio-hydrodynamic vortex need more attention.

The ABP model assumes that particles can absorb energy distributed over a
medium in which they move and store it in a depot in a way similar to animals
actively foraging for food over some space. Like animals, the particles must
expend energy in order to move through the medium, and they experience a
velocity-dependent dissipation. Also they expend energy to stay alive (similar
to metabolism). Furthermore, also like real animals, they are subject to
variable (random) forces, or ``noise''. A confining potential is
added in order that the particles are confined (or swarm). The model is
therefore well motivated from a biological point of view. Interestingly, in
addition to swarming, it predicts two basic motions that we designate the
noisy fixed point and the equally probable, symmetric limit cycle
pair. Similar motions are commonly observed in a variety of animals ranging in
size and complexity from bacteria to birds and fish. Moreover, the ABP model
has taught us that these two canonical particle motions arise from a minimal
set of conditions. The particles must exhibit a mean self-propelling velocity
and they must be confined either by an external potential or by the mean field
that arises from an interparticle attractive interaction.  Our experiments
with small densities of {\em Daphnia} demonstrate the two basic
motions. Moreover, the two limit cycles form the skeleton from which the
vortex is built after some symmetry-breaking process eliminates one in favor
of the other. At large particle densities the APB model was extended to
include symmetry breaking by Oseen hydrodynamic flow in one instance and by an
interparticle avoidance process, modeled by including a short range repulsive
potential, in another. Both processes lead to vortex formations, similar to
those observed with high density {\em Daphnia} swarms, but it remains to
investigate which one better describes the experimental observations.

The model described by the RWT has also taught us something
significant. Without including the peaked DTA, that signifies a statistical
bias in favor of forward motion, in the model, no limit cycle motions of the
walkers develop for any value of light intensity. With a uniform distribution
of turning angles (uncorrelated Brownian walkers), only noisy fixed point
motion is possible. Thus short-range temporal correlations, here indicated by
the twin peaks in the DTA, are necessary for rotational motions about the
central attractant. Within the framework of the RWT, we can now list four
minimal conditions necessary to form vortices: i) a self-propelling mean
velocity, ii) a statistical bias in favor of forward motion, iii) a central
attractant, and iv) a symmetry breaking process that tends to align local
velocities. The bimodal DTA, necessary for rotational motion, might be an
essential ingredient of the collective motions of large colonies of creatures
and thus might be a general feature detectable in other animals as well. Why
the natural movement of plankton follows such a distribution with maxima
located at approximately $\pm$35{\tc\symbol{176}} (true for both our {\em
Daphnia} and ocean-going zoo plankton) and how the animals actually choose to
turn either left or right at the end of each hop are unanswered questions.
But it is intriguing to speculate that these preferred turning angles may
somehow enhance fitness, for example by possibly maximizing the acquisition of
energy when food is distributed in patches. Indeed, we have calculated the
diffusion coefficients for a swarm of random walkers using the experimentally
determined bimodal DTA and found that, for parameters appropriate to the {\em
Daphnia}y, diffusion is considerably enhanced over what one would obtain for a
walker with purely uncorrelated motions. This strongly suggests that motions
characterized by a preferred turning angle confer an evolutionary advantage on
foraging animals.

\section*{Acknowledgments}

Supported by the Collaborative Research Center ``Complex Nonlinear Processes''
of the German Science Foundation, DFG-SFB555 (UE, WE, LSG) and the U.S.
Office of Naval Research, Physical Sciences Division (FM). AO gratefully
acknowledges financial support from a Feodor-Lynen Fellowship sponsored by the
Alexander von Humboldt Foundation. FM is grateful to the US-ONR and to the
Alexander von Humboldt Foundation for continuing support.  AO and FM also wish
to acknowledge Winfried Lampert for his patient tutorials and continuing
encouragement throughout the course of this work. We are also grateful to J.
Rudi Strickler an Ai Nihongi for stimulating discussions as well as their kind
hospitality and for allowing us to further explore the motions of {\em
  Daphnia} in his laboratory. We thank David Russell, Beatrix Beisner, Allan
Tessier and Lon Wilkens for valuable suggestions concerning zooplankton
culture and behavior and help with the experimental set-up. We are indebted to
Gabor Bal{\'a}zsi and Daniel Pflugfelder for some simulations and for initial
help with the RWT. UE and LSG acknowledge the fruitful discussions with Igor
M. Sokolov and Niko Komin concerning RWT and anomalous diffusion.

% \appendix

% \section{}
% \label{}

% Bibliographic references with the natbib package:
% Parenthetical: \citep{Bai92} produces (Bailyn 1992).
% Textual: \citet{Bai95} produces Bailyn et al. (1995).
% An affix and part of a reference:
%   \citep[e.g.][Ch. 2]{Bar76}
%   produces (e.g. Barnes et al. 1976, Ch. 2).

% \begin{thebibliography}{}

% % \bibitem[Names(Year)]{label} or \bibitem[Names(Year)Long names]{label}.
% % (\harvarditem{Name}{Year}{label} is also supported.)
% % Text of bibliographic item
  
% \bibitem[]{}

% \end{thebibliography}
\bibliographystyle{elsart-harv}
\bibliography{anke,allgemein,ameisen,bakterien,brown,cells,coherent,daphnia,erdmann,gbt,polymer,schleimpilze}

\end{document}